\documentclass[prd, twocolumn,nofootinbib,superscriptaddress]{revtex4}

\usepackage{subfigure}
\usepackage{diagbox}
\usepackage{multirow}
\usepackage{epsfig,latexsym,cancel,amssymb,amsmath,verbatim,mathrsfs}
\usepackage{color,xcolor,graphicx}

\usepackage{hyperref}
\hypersetup{
	colorlinks,
	linkcolor={red!70!black},
	citecolor={blue!80!black},
	urlcolor={blue!80!black},
	bookmarksopen=true
}
\newcommand{\beq}{\begin{equation}}
\newcommand{\eeq}{\end{equation}}
\newcommand{\bea}{\begin{eqnarray}}
\newcommand{\eea}{\end{eqnarray}}

\definecolor{darkgreen}{HTML}{228B22}

\allowdisplaybreaks[4]

\begin{document}

\title{ Scalar dark matter and Muon $g-2$ in a $U(1)_{L_{\mu}-L_{\tau}}$ model }
\author{\textsc{XinXin Qi}}
\author{\textsc{AiGeng Yang}}
\affiliation{Institute of Theoretical Physics, School of Physics, Dalian University of Technology, \\ No.2 Linggong Road, Dalian, Liaoning, 116024, People’s Republic of China}
\author{\textsc{Wei Liu}}
\affiliation{
Department of Applied Physics, Nanjing University of Science and Technology, \\ Nanjing 210094, People’s Republic of China}
\author{\textsc{Hao Sun}}
\email{haosun@dlut.edu.cn}
\affiliation{Institute of Theoretical Physics, School of Physics, Dalian University of Technology, \\ No.2 Linggong Road, Dalian, Liaoning, 116024, People’s Republic of China}

\begin{abstract}
We consider a simple scalar dark matter model within the frame of gauged $L_{\mu}-L_{\tau}$ symmetry.
A gauge boson $Z'$ as well as two scalar fields $S$ and $\Phi$ are introduced to the Standard Model (SM).
$S$ and $\Phi$ are SM singlet but both with $U(1)_{L_{\mu}-L_{\tau}}$ charge.
The real component and imaginary component of $S$ can acquire different masses after spontaneously symmetry breaking,
and the lighter one can play the role of dark matter which is stabilized by the residual $Z_2$ symmetry.
A viable parameter space is considered to discuss the possibility of light dark matter as well as co-annihilation case,
and we present current $(g-2)_{\mu}$ anomaly, Higgs invisible decay, dark matter relic density 
as well as direct detection constriants on the parameter space.
\end{abstract}
\maketitle

\section{Introduction}\label{sec:intro}

Recently, the Fermi-lab reported a new measurement
from the Muon $g-2$ experiment \cite{Abi:2021gix, Albahri:2021kmg, Albahri:2021ixb}, the result is:
\begin{eqnarray}
  a_{\mu}^{\rm exp}=116592061(41) \times 10^{-11} ,
\end{eqnarray}
compared with the SM prediction
\begin{eqnarray}
  a_{\mu}^{\rm SM}=116591810(43) \times 10^{-11} .
\end{eqnarray}
The observed $4.2\sigma$ discrepancy raises a big challenge to the SM and can give us new hints to the new physics.
Models related with $(g-2)_{\mu}$ anomaly can be found 
in \cite{Kawamura:2020qxo,Athron:2021iuf,Baltz:2001ts,Everett:2001tq,Choudhury:2001ad,Cheung:2001ip,Calmet:2001si,Xiong:2001rt,Cox:2021gqq,Chakrabarty:2018qtt,Chakrabarty:2020jro}.
Among these models, the gauged $U(1)_{L_{\mu}-L_{\tau}}$ has been discussed for a long time for its simplicity
since anomaly cancellation can be accomplished without introducing new fermions in such model.
Discussion about models for a new gauge boson ($Z'$) can be found in \cite{Foot:1994vd, Altmannshofer:2016brv,Foldenauer:2016rpi,Baek:2001kca,He:1990pn,He:1991qd,Belanger:2015nma,Heeck:2011wj},
where the $(g-2)_\mu$ anomaly can be naturally explained by the one loop contribution of $Z'$,
and the new gauge boson mass is limited to be not too heavy.
Searches for $Z^\prime$ at experiment can give the most stringent limit on the gauge boson mass and coupling constant,
and a viable parameter space with $Z'$ boson mass at MeV scale has been well studied,
which also satisfies current experiment constraints \cite{TheBABAR:2016rlg, Mishra:1991bv, Kaneta:2016uyt,Chun:2018ibr}.

The gauged $U(1)_{L_{\mu}-L_{\tau}}$ symmetry gives possible direction for the extension of SM,
and one can introduce new extra particles with $U(1)_{L_{\mu}-L_{\tau}}$ charge to the model to explain dark matter problem,
which is also a huge challenge to the SM. According to the astronomical observation,
our universe is not just composed of SM particles, but also dark matter as well as dark energy \cite{Komatsu:2010fb}.
Dark matter particles are assumed to be electrically neutral, colorless and stable on cosmological scales.
To explain the observed relic density of dark matter,
thermal Freeze-out \cite{Chiu:1966kg} and Freeze-in \cite{Hall:2009bx} mechanism have been put forward.
Aside from dark matter annihilating into SM particles,
other processes such as semi-annihilation \cite{Belanger:2014bga} and co-annihilation \cite{Baker:2015qna},
can also contribute to the relic density of dark matter.
New fermions as dark matter candidate in a gauged $L_{\mu}-L_{\tau}$ model can be found in \cite{Foldenauer:2018zrz, Altmannshofer:2016jzy},
while scalar dark matter models have been discussed in \cite{Biswas:2016yjr, Das:2021zea},
where the current relic density is obtained via Freeze-in mechanism.
Among these models, dark matter particles are almost stabilized by $U(1)_{L_{\mu}-L_{\tau}}$ symmetry.
Besides these models, one can also consider a type of so-called darkon DM matter model \cite{Cheung:2014ppa, He:2008qm}.
The darkon can play the role of dark matter via its lighter component
after spontaneously symmetry breaking in the case of $U(1)$ extension of the SM \cite{Chiang:2012qa,Farzan:2012hh,Farzan:2014foo}.
Particularly, one can have co-annihilation contribution when the real component and imaginary component of darkon are nearly degenerate.

In this article, we consider a simple scalar dark matter model within the frame of the gauged $L_{\mu}-L_{\tau}$ symmetry.
We introduce a new gauge boson $Z'$ and two scalar fields $S$ and $\Phi$ to the SM,
$S$ and $\Phi$ are singlet in the SM but carry $U(1)_{L_{\mu}-L_{\tau}}$ charge.
The scalar $\Phi$ will spontaneously breaking so that the new gauge boson $Z'$ acquires mass,
while the lighter component of $S$ will play the role of dark matter in our model after spontaneously breaking,
which will be stabilized with the residual $Z_2$ symmetry.
We focus on the $(g-2)_{\mu}$ anomaly problem and scalar dark matter in this work,
and discussion about neutrino mass problem in a similar gauged ${L_{\mu}-L_{\tau}}$ model can be found in \cite{Biswas:2016yan},
where the scalar dark matter in this model is stabilized by $U(1)_{L_{\mu}-L_{\tau}}$ symmetry.
We consider the possiblity of light dark matter as well as co-annihilation case in our model 
and we show the Higgs invisble decay, relic density constraint as well as direct detection constraint on a viable parameter space.

This article is arranged as followed:
We give the description of the scalar dark matter model in section \ref{sec:2}.
We consider the $(g-2)_{\mu}$ anomaly constraint on the parameter space in section \ref{sec:3}.
We discuss Higgs invisible decay, relic density as well as direct detection constraint
on the chosen parameter space separately in \ref{sec3.1}, \ref{sec3.2} and \ref{sec3.3}.
We give a summary in the last section \ref{sec:so}.

\section{model description}\label{sec:2}

We discuss a scalar dark matter model based on the $U(1)_{L_{\mu}-L_{\tau}}$ extension of the SM.
A new gauge boson $Z'$ as well as two SM singlet scalar $\Phi$ and $S$ are introduced to the standard model.
The field $H$ gives the non-zero vaccum expection value (vev) like in the SM,
while $\Phi$ develops a vev to break the $U(1)_{L_{\mu}-L_{\tau}}$ symmetry.
The field $S$ has zero vev and plays the role of scalar dark matter via its lighter component.
$Z'$ will acquire mass after symmetry breaking spontaneously.
We can assume $S$ and $\Phi$ carry opposite $U(1)_{L_{\mu}-L_{\tau}}$ charge
so that we have $\lambda_{ds}(S^2\Phi^2 +{S^\dagger}^2{\Phi^\dagger}^2)$ in the lagrangian,
which is essential because it triggers the $U(1)_{L_{\mu}-L_{\tau}} \to Z_2$ spontaneously breakdown as soon as $\Phi$ acquires a vev.

The additional part of the fermion lagrangian is given by:
\begin{align} \nonumber
  {\cal L}_{\rm fermion} &= g_p Z^{\prime\sigma} (\bar{\mu} \gamma^\sigma \mu - \bar{\tau} \gamma^\sigma \tau \\
  &+ \bar{v}_\mu \gamma^\sigma P_L v_\mu - \bar{v}_\tau \gamma^\sigma P_L v_\tau).
\end{align}
The scalar part with dark matter is given by:
 \begin{eqnarray}
  {\cal L}_{\rm scalar}= |D_\mu \Phi |^2 + |D_\mu H|^2 + |D_\mu S|^2 - {\cal V}(H, \Phi, S)
\end{eqnarray}
with
\begin{align} \nonumber
&  D_\mu\Phi = \partial_\mu \Phi - ig_p q_1 A'_\mu\Phi , \\
&  D_\mu H = \partial_\mu H - ig_L W_\mu H -\frac{i}{2} g_Y A_\mu H, \\\nonumber
&  D_\mu S = \partial_\mu S - ig_p q_2 A'_\mu S ,
\end{align}
where $q_1$ and $q_2$ are the $U(1)_{L_{\mu}-L_{\tau}}$ charge of $\Phi$ and $S$, $g_p$ is the $ U(1)_{L_{\mu}-L_{\tau}}$ coupling constant.
For simplicity, we assume $q_1=2$ and $q_2=-2$ as we considered above.
The potential term ${\cal V}(H, \Phi, S)$ is given by:
\begin{align} \nonumber
 {\cal V}(H, \Phi, S) & = \mu |H|^2 + \lambda_H |H|^4 +\mu_1 |\Phi |^2 + \lambda_p |\Phi |^4 \\ \nonumber
                      & - \frac{1}{2}m_0^2 |S|^2 + \frac{\lambda_s}{4} |S|^4 + \lambda_{Hp} |H|^2 |\Phi |^2 \\ \nonumber
			    & + \lambda_{Hs} |H|^2 |S|^2 + \lambda_{SP} |S|^2 |\Phi |^2 \\
			    & + \frac{ \lambda_{ds}}{2} (S^2 \Phi ^2 +{S^\dagger}^2 {\Phi^\dagger}^2) .
\end{align}
where $\mu <0$ and $\mu_1<0$ and $m_0^2>0$. We assume $H$ and $\Phi$ spontaneously breaking, and we have $H$ and $\Phi$ in unitary gauge form with:
\begin{eqnarray}
  H=  \left(
  \begin{array}{c}
    0\\
    \frac{v+h}{\sqrt{2}}
  \end{array}
      \right),
~~~\Phi = \frac{v_b +\phi}{\sqrt{2}}
\end{eqnarray}
where $v$ and $v_b$ are the vevs and we assume $v$ is the SM one equal 246 GeV.
What's more, $v_b =M_{Z_p}/2g_p$, where $M_{Z_p}$ is the new gauge boson $Z'$ mass.
Furthermore, we can write $S = S_R +i S_I$ in the form of real component and imaginary component, and we have:
\begin{eqnarray}
 m_R^2 &=& m_0^2+ \lambda_{SP}v_b^2 + \lambda_{Hs}v^2 + \lambda_{ds}v_b^2 \\
 m_I^2 &=& m_0^2+ \lambda_{SP}v_b^2 + \lambda_{Hs}v^2 - \lambda_{ds}v_b^2
 \label{eq8}
\end{eqnarray}
where $m_R^2$, $m_I^2$ correspond to the squared mass of $S_R$ and $S_I$.
The mass difference between $S_R$ and $S_I$ is determined by the sign of $\lambda_{ds}$.
We can take $\lambda_{ds}>0$ so that the lighter particle $S_I$ is the weakly interacting massive particle (WIMP) dark matter.
The result will be the same if we take $\lambda_{ds}<0$ while $S_R$ acts as the dark matter.

The mass matrix related with two Higgs is given by the following, after spontaneously breaking:
\begin{eqnarray}
 \label{ma}
 {\cal M}= \left(
   \begin{array}{cc}
   2\lambda_H v^2 &\lambda_{Hp} vv_b\\
   \lambda_{Hp}vv_b& 2\lambda_p v^2_b\\
   \end{array}
   \right) .
\end{eqnarray}
The mass matrix eigenvalue can be analytically expressed, and the result is given by:
\begin{eqnarray}
  m_1^2 &=& \lambda_H v^2 + \lambda_p v_b^2 -\sqrt{(\lambda_H v^2-\lambda_p v_b^2)^2 + \lambda_{Hp}^2 v_b^2 v^2} ~~~~\\
  m_2^2 &=& \lambda_H v^2 +\lambda_p v_b^2 + \sqrt{(\lambda_H v^2-\lambda_p v_b^2)^2 + \lambda_{Hp}^2 v_b^2 v^2} .~~~~~~
\end{eqnarray}
The gauge eigenstate and mass eignestate is related with a mixing angle $\theta$ which can be determined
\begin{eqnarray}
   \sin 2\theta = \frac{\lambda_{Hp} v_b v}{\sqrt{(\lambda_H v^2 -\lambda_p v^2_b)^2 + \lambda_{Hp}^2 v_b^2 v^2}} .
\end{eqnarray}
We consider the lighter Higgs is just the SM Higgs observed at the LHC with $m_1=125$GeV and $m_2$ being another Higgs mass in this work.
We use $h_1$ to represent the SM Higgs and $h_2$ to represent another Higgs for convenience.
The relation between the mass eigenstates and mass eigenstates can be given by:
 \begin{equation}
   \left(
 \begin{array}{c}
  h_1 \\
  h_2 \\
  \end{array}
  \right)=
  \left(
  \begin{array}{cc}
  \cos\theta & -\sin\theta \\
  \sin\theta &  \cos\theta \\
  \end{array}
  \right)
  \left(
  \begin{array}{c}
   h \\
  \phi \\
  \end{array}
  \right)
  \end{equation}
and we can express the couplings with two Higgs mass and mixing angle as followed:
 \begin{eqnarray}
&& \lambda_H = \frac{m_1^2}{4v^2}(1+\cos 2\theta)+\frac{m_2^2}{4v^2}(1-\cos 2\theta)  \\
&& \lambda_p = \frac{m_1^2}{4v_b^2}(1-\cos 2\theta)+\frac{m_2^2}{4v_b^2}(1+\cos 2\theta) \\
&& \lambda_{Hp} = \sin 2\theta \frac{m_2^2-m_1^2}{2vv_b} .
\label{eq15}
 \end{eqnarray}

To make sure the model perturbative, contribution from loop correction should be smaller 
than tree level values, such constraint can be ensured when \cite{Chakrabarty:2015yia}
\begin{align} \nonumber
& |\lambda_{H}| < 4\pi,\ |\lambda_{Hp}| < 4\pi,\ |\lambda_{p}| < 4\pi, |\lambda_{Hs}| < 4\pi,\\
& |\lambda_{ds}| < 4\pi,\ |\lambda_{SP}|<4 \pi .
\end{align}
What's more, to obtain a stable vaccum, the quartic couplings apprearing in the lagrangian should be constrained,
we consider the necessary and sufficient conditions as followed \cite{Kannike:2016fmd, Biswas:2016ewm}:
\begin{align} \nonumber
&  \lambda_H \geq 0, \lambda_s \geq 0, \lambda_p \geq 0, \lambda_{Hs} \geq -2 \sqrt{\lambda_H\lambda_s}, \\ \nonumber
&\lambda_{Hp} \geq -2 \sqrt{\lambda_H\lambda_p},\ \lambda_{SP} \geq -2 \sqrt{\lambda_P\lambda_s}, \\ \nonumber
&  \sqrt{\lambda_{Hs} +2 \sqrt{\lambda_H\lambda_s}}\sqrt{\lambda_{Hp} +2 \sqrt{\lambda_H\lambda_p}}\sqrt{\lambda_{SP} +2 \sqrt{\lambda_p\lambda_s}} \\ \nonumber
&  +2\sqrt{\lambda_H\lambda_s\lambda_p}+\lambda_{Hp}\sqrt{\lambda_s}+\lambda_{Hs}\sqrt{\lambda_p} \\
&  +\lambda_{SP}\sqrt{\lambda_H} \geq 0 .
\end{align}
In this work, we choose the following parameters as the inputs with
\begin{eqnarray}
  m_2,\ M_{Z_{p}},\ g_p,\ \lambda_{s},\ \lambda_{ds},\ \lambda_{SP},\ \lambda_{Hs},\ \sin\theta, \ m_0.
\end{eqnarray}

\section{Parameter constraints on $Z'$}\label{sec:3}

\begin{figure}[htbp]
  \includegraphics[scale=0.4]{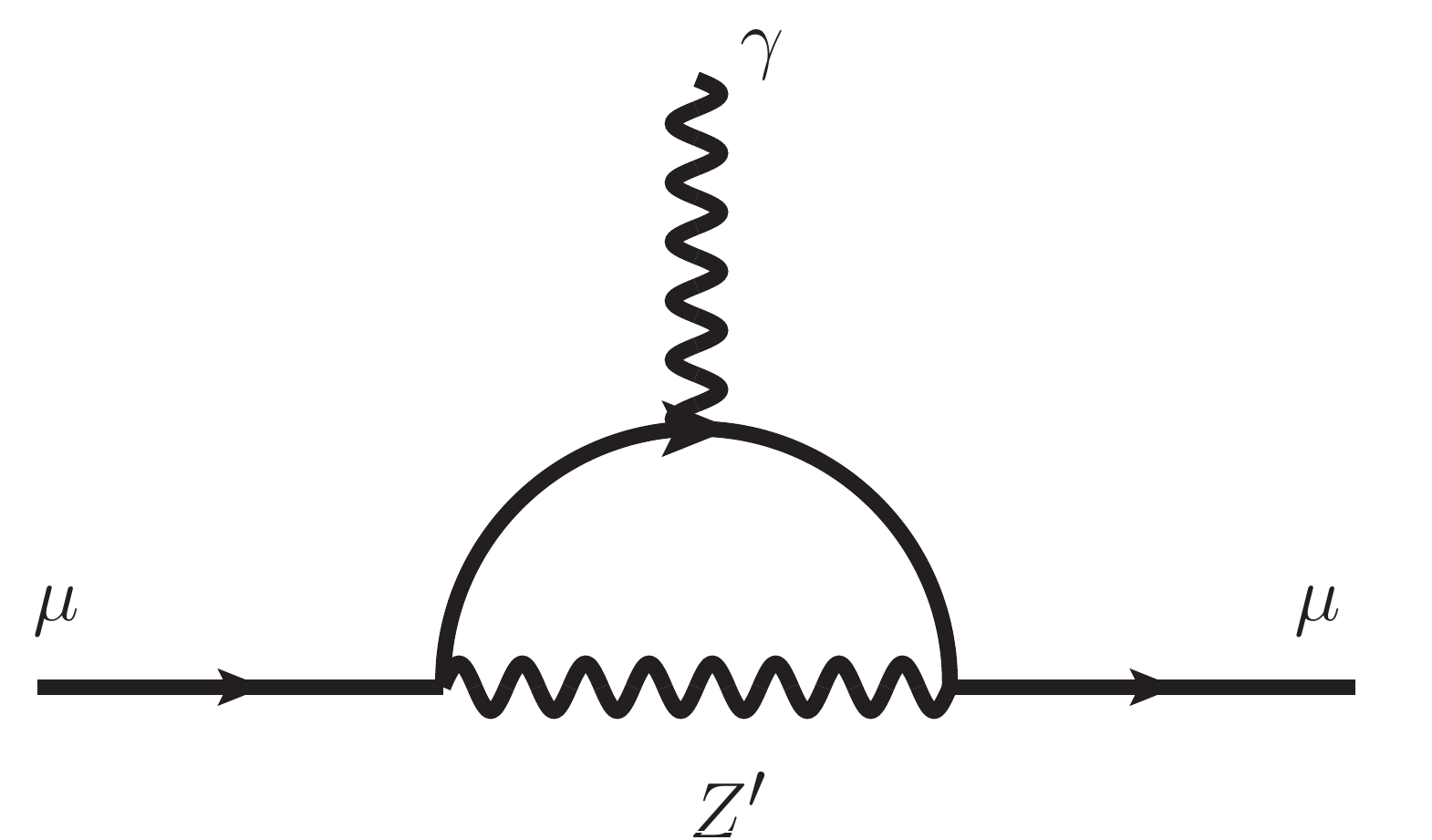}
  \caption{One-loop contribution of $Z'$ to $(g-2)_{\mu}$ anomaly.}
  \label{Fig:fig1}
\end{figure}
Before discussion, we should consider the $(g-2)_{\mu}$
constraints on $ M_{Z_p}-g_p$ plane, since one prospect to
introduce $U(1)_{L _{\mu}- L_{\tau}}$ symmetry is to explain the $(g-2)_{\mu}$ anomaly.
Contribution of the new gauge boson $Z'$ to the muon anomalous magnetic momentum is shown in Figure \ref{Fig:fig1}.
The analytical expression for $\Delta a_\mu$ is
\begin{eqnarray}
 \Delta a_\mu = \frac{g_p^2 m_{\mu}^2}{8\pi^2} \int_{0}^{1} \frac{2x^2(1-x)}{x^2 m_{\mu}^2+(1-x)M_{Z_{p}}^2} dx
 \label{g-2}
\end{eqnarray}
where $m_{\mu}$ is the muon mass.

Searches for new gauge bosons at experiment has already given the most stringent constraint on the parameter space of $M_{Z_p}-g_p$.
At colliders, the $Z'$ gauge boson can be produced via $e^+ e^- \to \mu^+ \mu^- Z'$
with the subsequent decay of $Z'\to\mu\mu$. Such search can be found in Babar \cite{TheBABAR:2016rlg}
and give us the possible bound on $M_{Z_p}-g_p$ plane.
What's more, neutrino expriments give us new clues to constrain the parameter space.
The Borexino data related with the scattering of low energy solar neutrinos \cite{Agostini:2017ixy, Abdullah:2018ykz}
can provide the most stringent constraint on the low $M_{Z_p}$ and low $g_p$ region.
Another constraint is from CCFR collaboration \cite{Altmannshofer:2014pba}, obtained via the neutrino trident production
which is related with the process $\nu N \to \nu N \mu^+ \mu^-$,
where a muon neutrino scattered off of a nucleus producing a $\mu^+ \mu^-$ pair.
Such process will be enhanced due to the existence of $Z'$ compared with SM case,
which gives strong bounds on the possible $Z'$ contribution and constrain the $M_{Z_p}-g_p$ parameter space accordingly.
  \begin{figure}[htbp]
  \vspace{0.2cm}
  \hspace{-0.8cm}
  \includegraphics[scale=0.65]{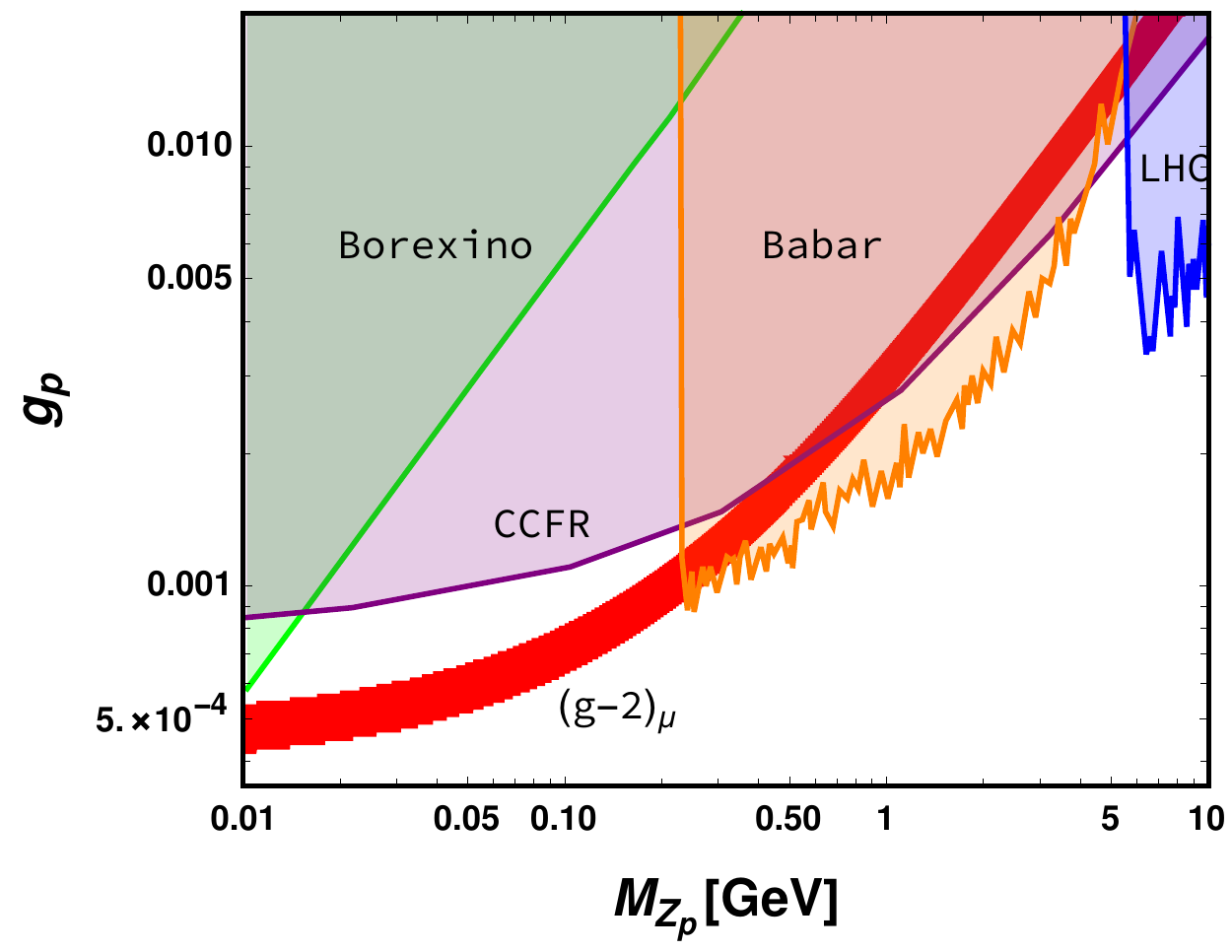}
  \caption{
  The allowed region that can explain the $(g-2)_{\mu}$ anomaly and the exclued region on $M_{Z_p}-g_p$ from different experiments:
  the green (orange, blue, purple) region is excluded by the Borexino (BABAR, LHC, CCFC).
  The red region is the parameter satisfying $(g-2)_{\mu}$ anomaly.
  }
 \label{Fig:fig2}
\end{figure}
In our paper, we focus on the low $M_{Z_p}$ region.
The combined experiment as well as $(g-2)_{\mu}$ constraints on the $M_{Z_p}-g_p$ is given in Figure \ref{Fig:fig2}.
According to Figure \ref{Fig:fig2}, CCFR gives the most stringent constraint on the parameter space we interest at low $M_{Z_p}$ region.
The Borexino limits become relavant at smaller $M_{Z_p}$ and $g_p$.
For $M_{Z_p} \subseteq [0.2,5]$ GeV, BABAR data plays an important role to constrain $M_{Z_p}$ and $g_p$.

\begin{figure}[htbp]
  \hspace{-0.8cm}
  \includegraphics[scale=0.32]{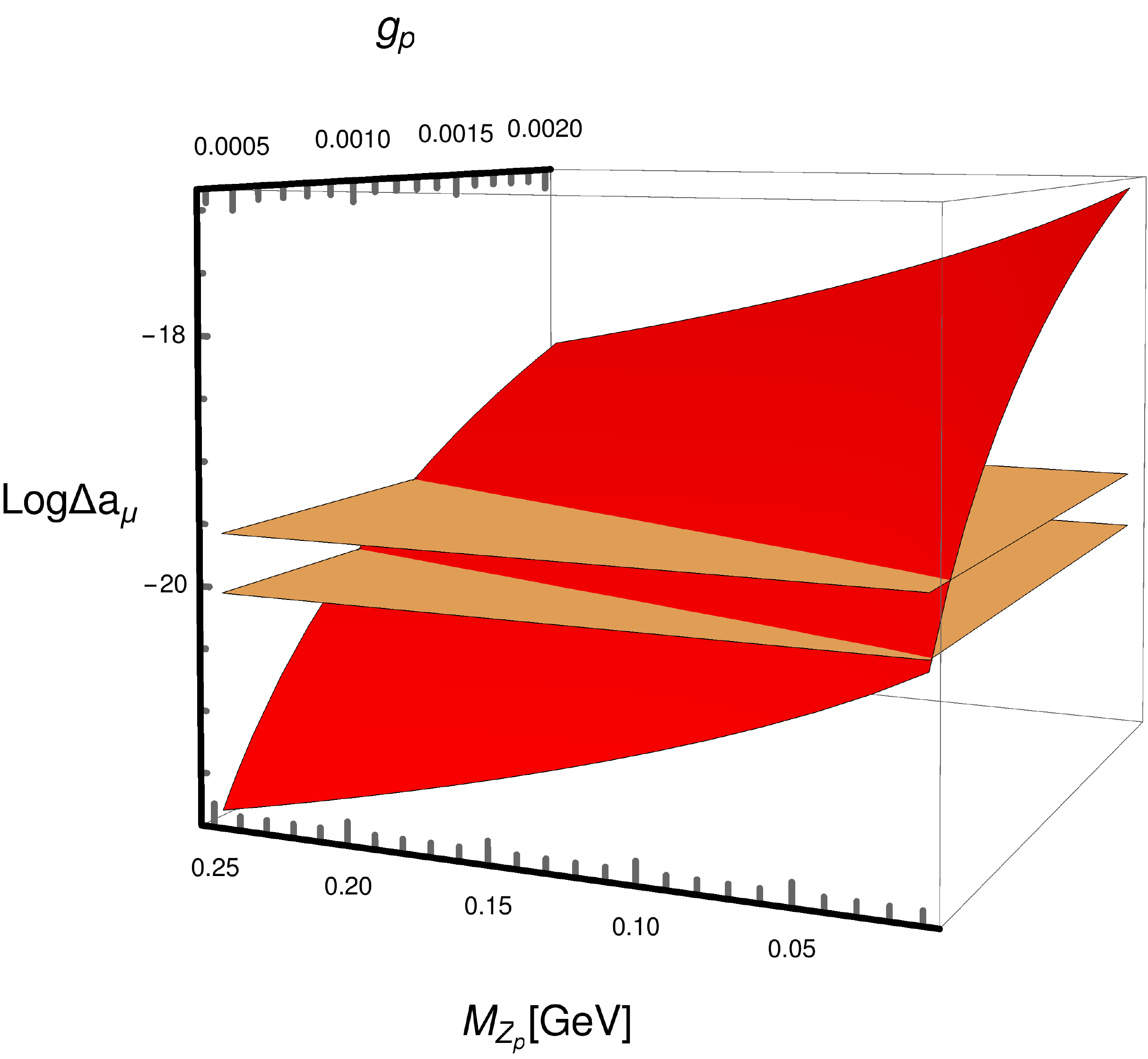}
  \caption{
$M_{Z_p}-g_p$ plane to explain $(g-2)_{\mu}$ anomaly with $M_{Z_p} \subseteq [0.01,0.25]$ GeV.
The red region is Log result of Eq.(\ref{g-2}) and the two yellow regions are the latest experiment result of $(g-2)_{\mu}$.
The overlap part of the three regions is the viable parameter space to explain $(g-2)_{\mu}$ anomaly within experiment constraints.}
  \label{Fig:fig3}
\end{figure}
To sum up, most of the parameter space to explain $(g-2)_{\mu}$ has been excluded by these experiments
but $M_{Z_p} \subseteq [0.01,0.25]$ GeV and $g_p$ at $10^{-3}$ level.
The viable parameter region that can explain the $(g-2)_{\mu}$ anomaly in the $M_{Z_p}-g_p$ plane
is shown in Figure \ref{Fig:fig3} accoding to the latest experiment value $\Delta a_\mu = 251(59) \times 10^{-11}$
of where the red plane is the result of $\rm Log\Delta a_{\mu}$ and the two yellow palnes 
are the low bound and top bound on $\rm Log\Delta a_{\mu}$ from the experiment. 
Intersection region of the three planes is the viable parameter space of $M_{Z_p}-g_p$ explaining $(g-2)_{\mu}$ anomaly, 
and we can give some pairs of $(M_{Z_p}, g_p)$ with $(0.01 \rm GeV, 4.18 \times 10^{-4})$, $(0.25 \rm GeV, 1.117 \times 10^{-3})$.

\section{Phenomenological study}

\subsection{Higgs invisible decay } \label{sec3.1}

In our model, we have a new gauge boson $Z'$ and two  scalar fields $S_I$ and $S_R$,
We assume all these particles masses are smaller than the SM Higgs mass
with $M_{Z_p} <1/2 m_1$, $m_I< 1/2 m_1$ and $m_R< 1/2 m_1$  so that
SM Higgs can decay to these particles which will contribute to the invisible Higgs decay width.
Current constraint on such invisible branching fraction is ${\rm Br}(h \to inv) <0.24$
accroding to the observations at the LHC for the SM Higgs \cite{Tanabashi:2018oca}, which means
\begin{eqnarray}
   \frac{\Gamma(h \to inv)}{\Gamma(h \to inv)+\Gamma(h \to {\rm SM})} <0.24 .
   \label{gam}
\end{eqnarray}
The decay widths related with Higgs invisible decay channel in our model are given as followed:
\begin{eqnarray} \nonumber
\Gamma_{h_1 \to S_I S_I} &=& \sqrt{m_1^2 (m_1^2-4 m_{I}^2)}/(8 \pi m_1 ^3) \\ \nonumber
                         &\times& (\cos\theta\lambda_{Hs}v+\lambda_{ds} \sin\theta v_b -\lambda_{SP} \sin\theta v_b)^2\\ \nonumber
\Gamma_{h_1 \to S_R S_R} &=& \sqrt{m_1^2 (m_1^2-4 m_{R}^2)}/(8 \pi m_1 ^3) \\  \nonumber
                         &\times& (\cos\theta\lambda_{Hs}v-\lambda_{ds} \sin\theta v_b -\lambda_{SP} \sin\theta v_b)^2\\ \nonumber
\Gamma_{h_1 \to Z' Z'}   &=& g_p^4\sqrt{m_1^2 (m_1^2-4 M_{Z_p}^2)} /(2 M_{Z_p}^4 \pi m_1 ^3) \\
                         &\times& (m_1^4-4m_1^2M_{Z_p}^2+ 12M_{Z_p}^4) \sin^2\theta v_b^2 .
\end{eqnarray}

\begin{figure}[htbp]
  \vspace{0.4cm}
  \hspace{-0.1cm}
  \includegraphics[scale=0.5]{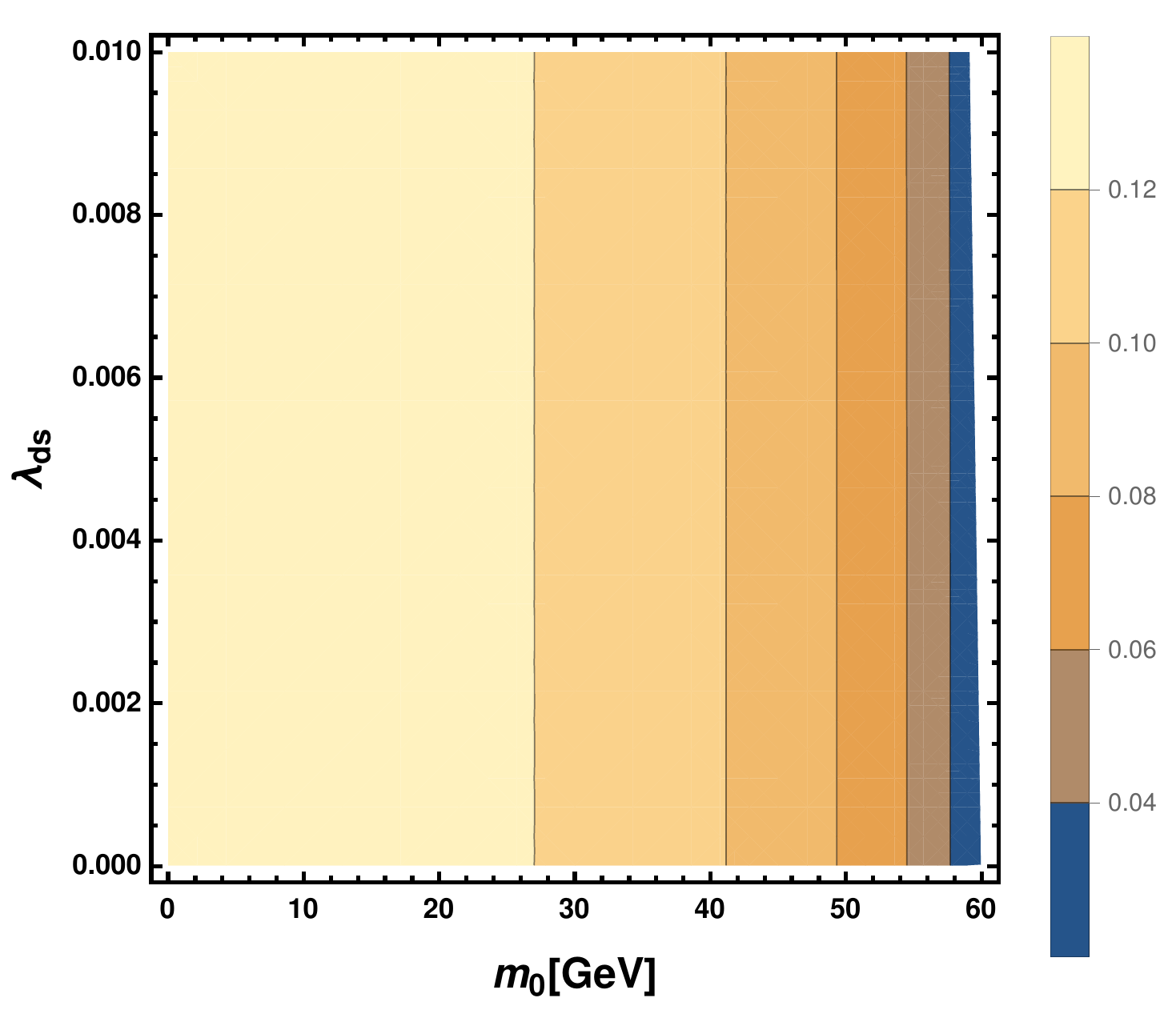}
  \caption{Contourplot of Higgs invisible decay width where the X-axis is $m_0$ 
  and Y-axis is $\lambda_{ds}$ when $\lambda_{Hs}=0.005$, $\lambda_{SP}=0.001$.}
  \label{Fig:fig4}
\end{figure}
In this part, we consider the Higgs invisible decay constraint on the chosen parameter space. 
As we can see in the following discussion, $\lambda_{ds}$ is limited stringetly under relic density constraint, 
while $\lambda_{Hs}$ and $\lambda_{SP}$ are more flexible. According to Eq.(\ref{gam}), 
the contribution of $\lambda_{Hs}$ to Higgs invisble decay width is positive while $\lambda_{SP}$ is negative. 
For simplicity, we can set $\lambda_{Hs}$ and $\lambda_{SP}$ to be some special value 
and give the result of the Higgs invisible decay width of our model to estimate the chosen parameter space.
In Figure \ref{Fig:fig4}, we fixed $\lambda_{Hs}=0.005$ and $\lambda_{SP}=0.001$
and set $\lambda_{ds} \subseteq [1 \times 10^{-5},0.01]$ and $ m_0 \subseteq [0,60] GeV$. 
According to Figure \ref{Fig:fig4}, the Higgs invisible decay width 
is much lower than current constraint in the case of $\lambda_{Hs}=0.005$ and $\lambda_{SP}=0.001$. 
As we discussed above, we can set $\lambda_{Hs}<0.005$ and $\lambda_{SP}>0.001$ 
to get a viable parameter space satisfying Higgs invisible decay width constraint.

\subsection{Relic density}\label{sec3.2}

In this part, we discuss the dark matter phenomenology in the model. The expression of relic abundance of dark matter can be given as followed:
\begin{eqnarray}
 \Omega h^2 = \frac{1.07\times 10^9 {\rm GeV}^{-1} }{{g*}^{1/2}M_{Pl}}\frac{1}{J(x_f)}
\end{eqnarray}
where $g*$ is the total number of effective relativistic degrees of freedom, $M_{Pl}=1.22\times 10^{19}$ GeV is the Planck mass, and $J(x_f)$ is given by:
\begin{eqnarray}
 J(x_f)=\int _{x_f}^\infty \frac{dx}{x^2} \langle \sigma {\rm v} \rangle (x) .
\end{eqnarray}
The freeze-out parameter $x_f$ in the integral is \cite{kolb1981early}:
\begin{eqnarray}
 x_f =\ln \frac{0.038 g M_{Pl} m_{\rm DM} \langle \sigma {\rm v} \rangle (x_f)}{( {g*} x_f )^{1/2}}
\end{eqnarray}
where $m_{\rm DM}$ is the dark matter mass. 
According to current experiment, the dark matter relic density is \cite{Aghanim:2018eyx}:
\begin{eqnarray}
  \Omega h^2 =0.1198 \pm 0.0012 .
\end{eqnarray}

Scalar dark matter in our model is stabilized by $Z_2$ symmetry after $U(1)_{L_{\mu}-L_{\tau}}$ symmetry spontaneously breaking.
It is worth stressed that another scalar $S_R$ can also play role of dark matter when $S_I$ and $S_R$ are nearly degenerate.
The dark matter relic density will not just be determined by $S_I$ but also $S_R$, which is so-called co-annihilation case \cite{Griest:1990kh}.
In the case we considered, the number density of $S_R$ will track $S_I$ number density during freeze-out,
when the relative mass splitting $\Delta$ is small compared to the freeze-out temperature, which is defined by:
 \begin{eqnarray}
 \Delta \equiv \frac{m_R-m_I}{m_I} .
 \end{eqnarray}
Concretely speaking, the relic density is calculated by solving the Boltzman equation,
which depends on the dimensionless variable $x=m_{\rm DM}/T$.
Freeze-out occurs at $x_F=m_{\rm DM}/T_F \approx 20\sim 30$ for cold,
non-relativistic dark matter, where $T_F$ is the freeze-out temperature.
For $\delta m = m_R-m_I $ much larger than $T_F$,
we have just dark matter $S_I$ annihilations freeze-out.
However, $S_R$ will be thermally accessible when $\delta m \approx T_F$.
Hence, the relative mass splitting $\Delta$ can give the upper bound
that $\Delta \sim x_F^{-1} \sim 0.03-0.05$ when we consider the co-annihilation case.
A more systematic estimate for the contribution of $\Delta$ to relic density can be found in \cite{Baker:2015qna}.
$\Delta$ in our model can be given as followed:
\begin{eqnarray}
 \Delta = \frac{m^2_R-m^2_I}{m_I(m_I+m_R)}=\frac{2\lambda_{ds}v^2_b}{m^2_I+m_Im_R}\label{24}.
\end{eqnarray}
According to Eq.(\ref{24}), co-annihilation becomes more significant in the model as long as $\lambda_{ds} \ll m_I$,
where $S_I$ and $S_R$ are kept in equilibrium via the interactions $S_I S_I \leftrightarrow S_R S_R$,
while in the light dark matter case the main contribution of relic density arising from the annihilation of $S_I$.
Before we consider the relic density numberically, we should stress that
co-annihilation process can be dominated during the evolution of relic denstiy in the heavy dark matter case,
but we want to consider a viable region where annihilation and co-annihilation can be both involved
which means the dark matter mass we considered should not be too heavy.
We choose a viable parameter space with $0\leqslant m_0\leqslant 60$ GeV
and dark matter particle mass will be constrained to be a few GeV to about 70 GeV.
Annihilation process and co-annihilation will be both involved within the chosen parameter space as we can see in the following discussion.

\begin{figure}[htbp]
\centering
  \includegraphics[height=2cm,width=2.5cm]{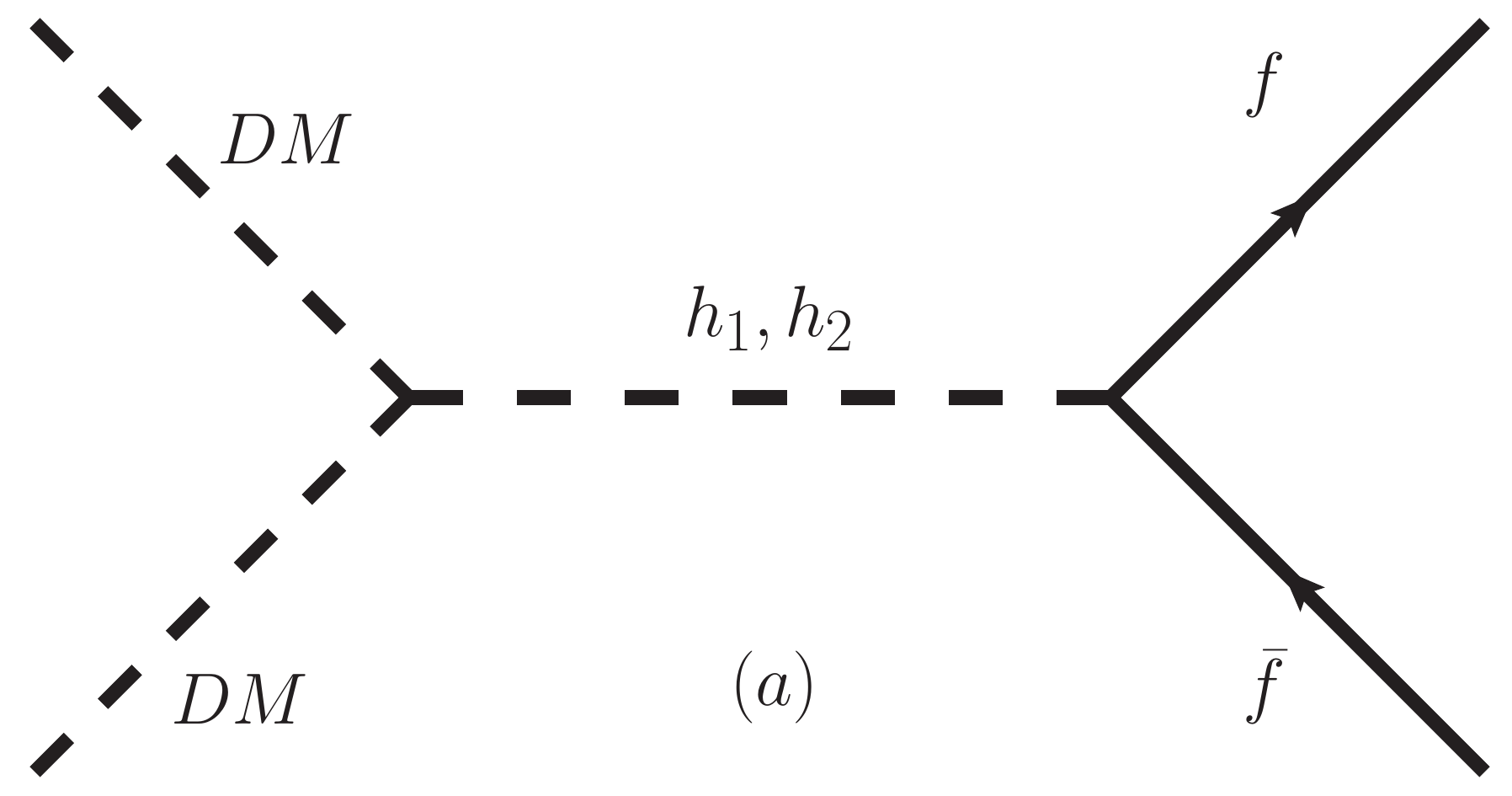}
\hspace{0.1cm}
  \includegraphics[height=2cm,width=2.5cm]{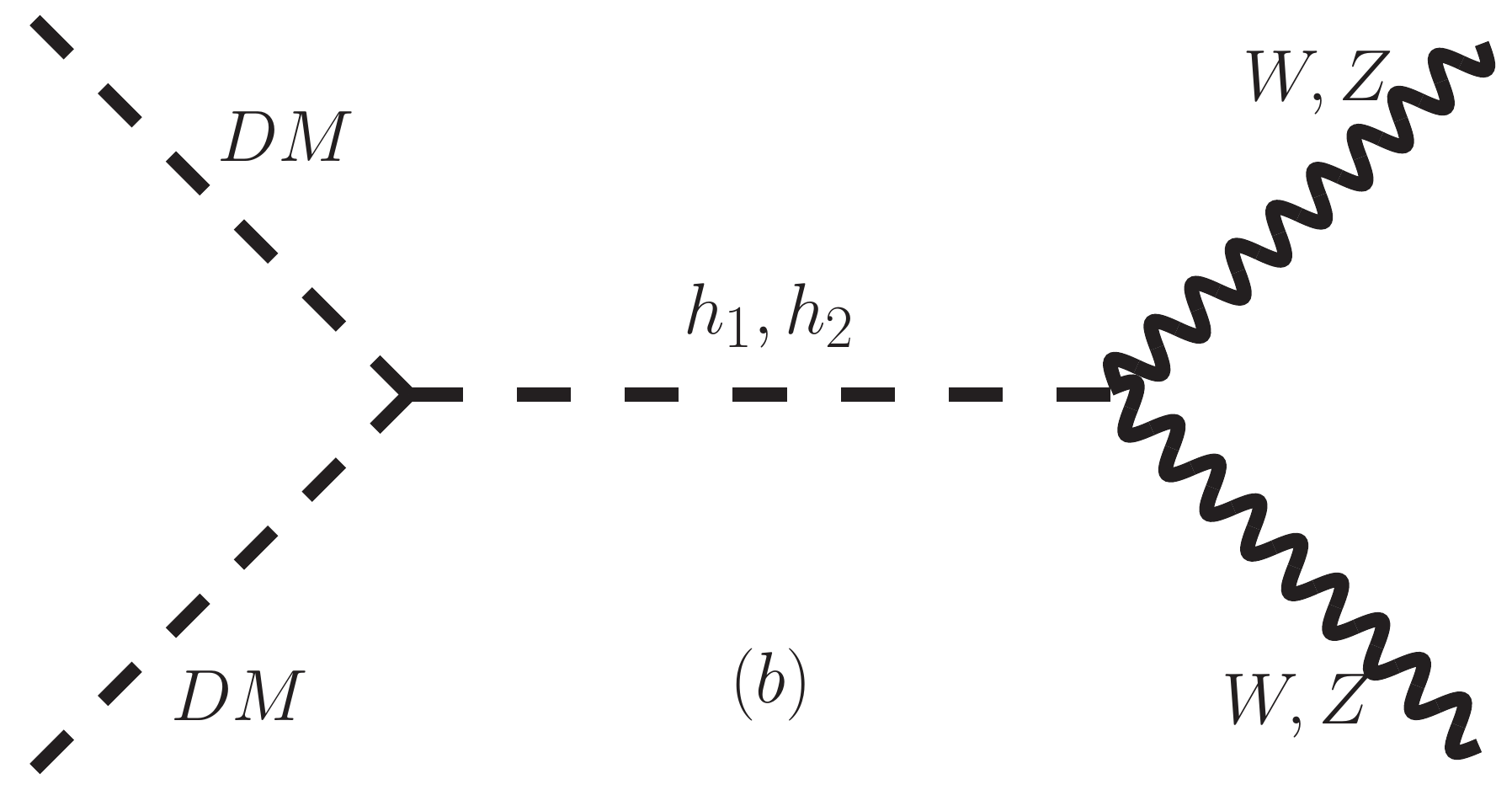}
\hspace{0.1cm}
  \includegraphics[height=2cm,width=2.5cm]{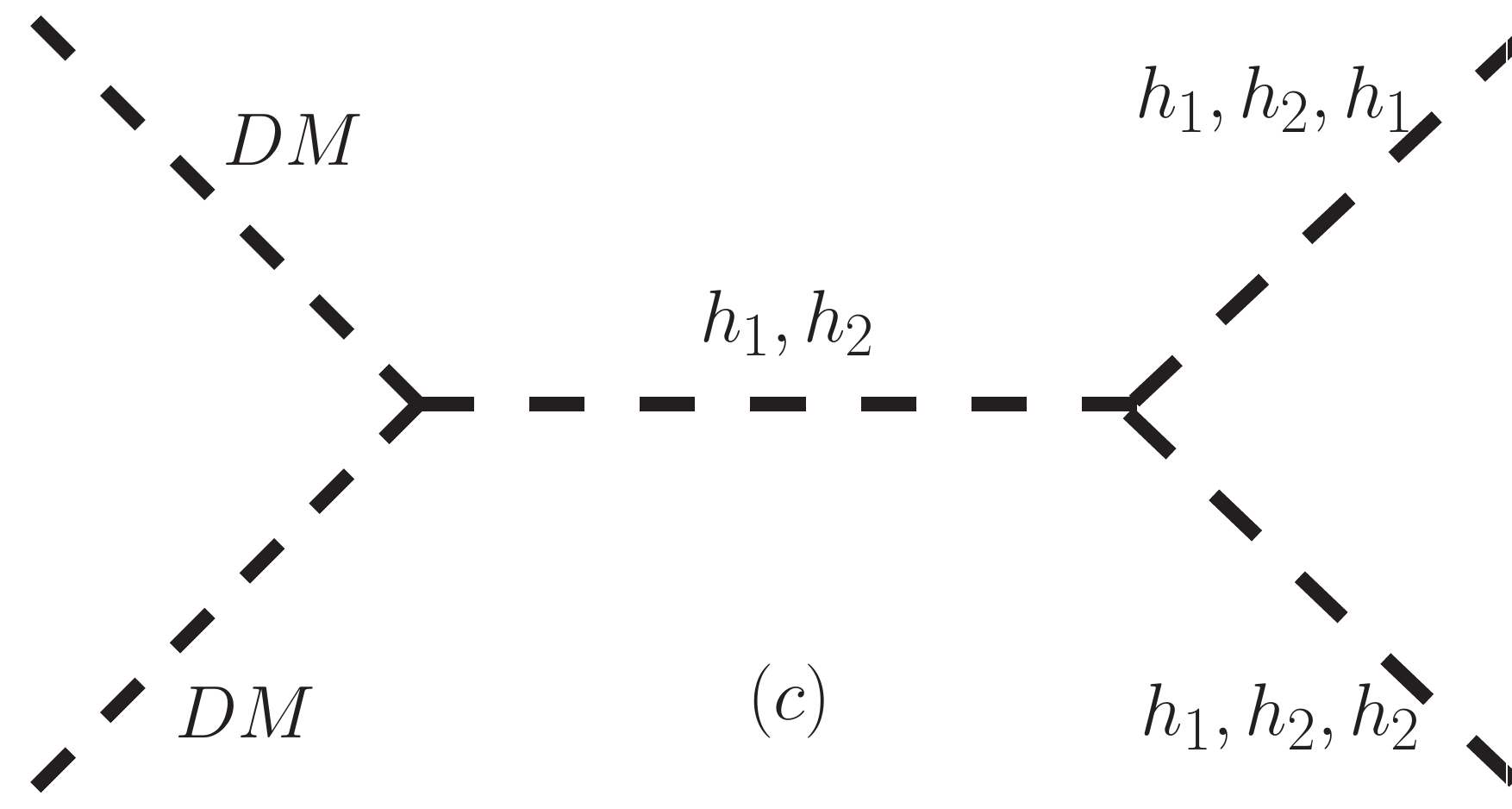}\\
\vspace{0.1cm}
\hspace{0.1cm}
  \includegraphics[height=2cm,width=2.5cm]{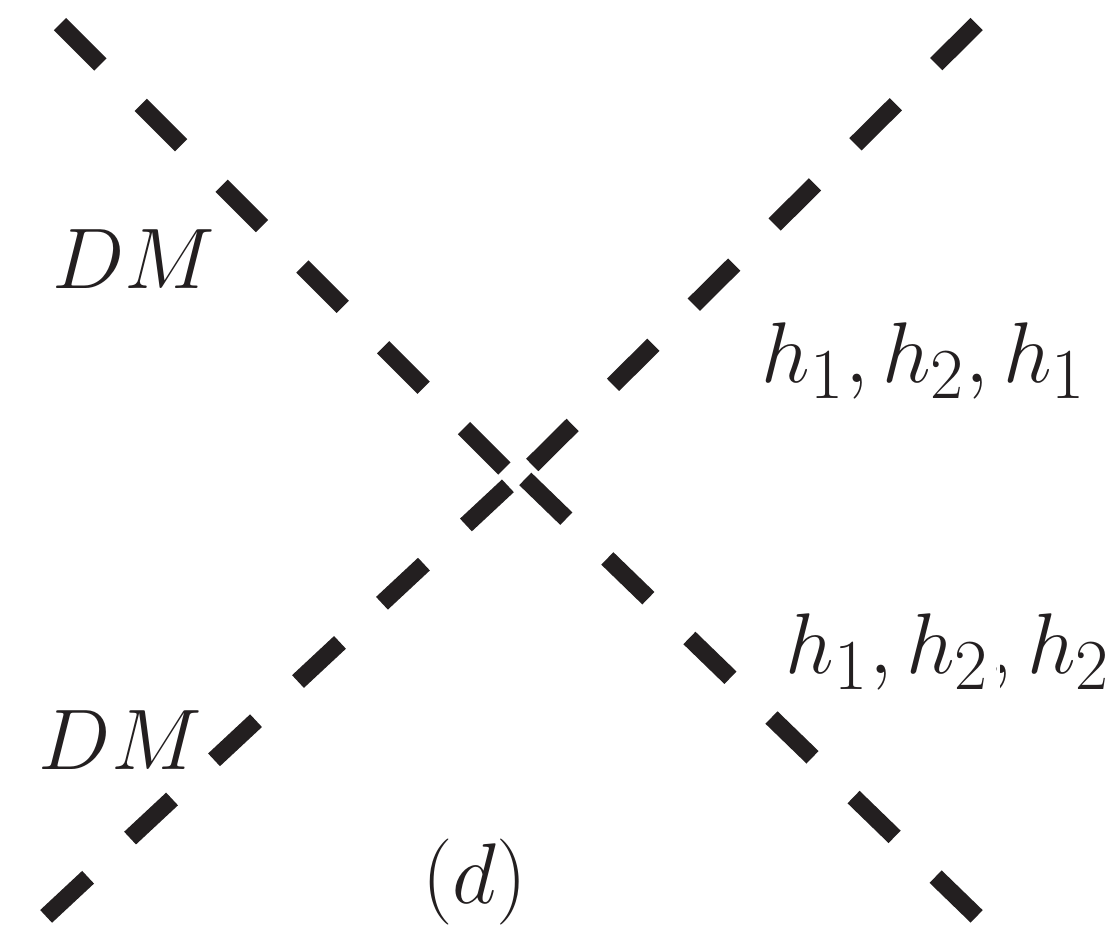}
\hspace{0.1cm}
  \includegraphics[height=2cm,width=2cm]{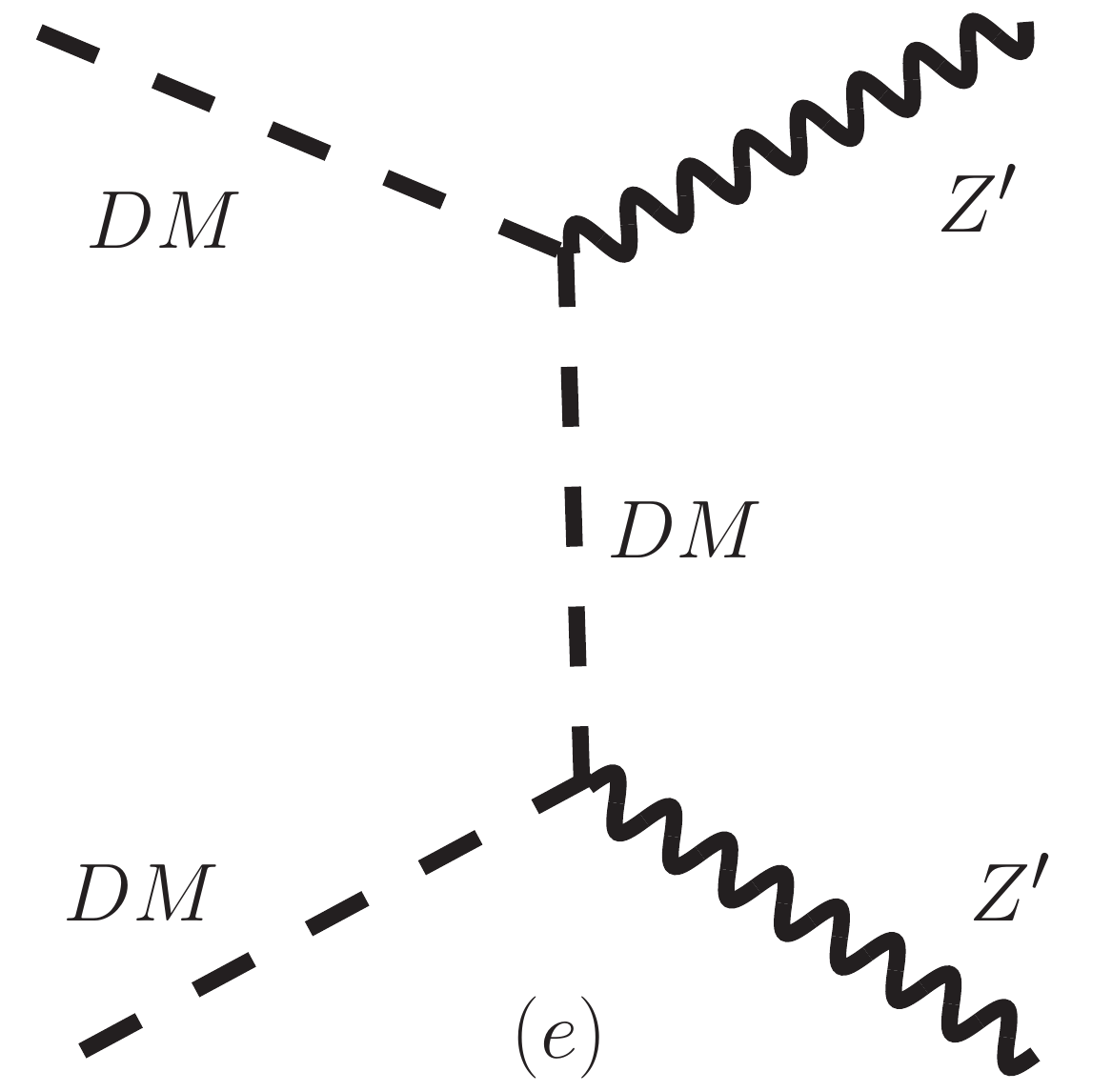}
\hspace{0.1cm}
  \includegraphics[height=2cm,width=2cm]{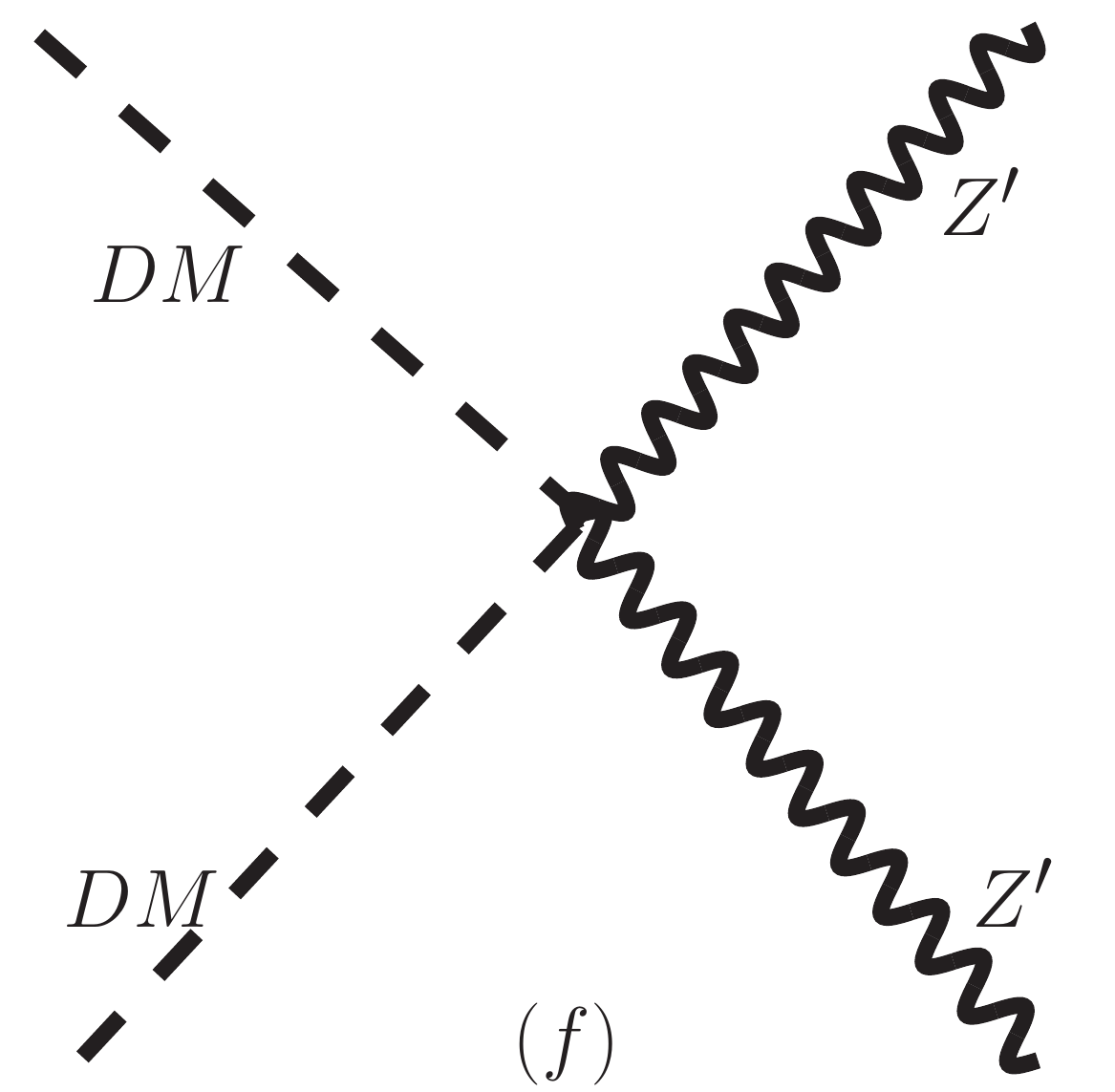}
  \caption{ Feynman diagrams related with dark matter relic density.}
 \label{Fig:fig5}
  \end{figure}

The Feynman diagrams relavant for the dark matter production are given in Figure \ref{Fig:fig5}.
According to Figure \ref{Fig:fig5}, scalar dark matter can annihilate into SM particles with Higgs-mediated interactions.
In the case of $2m_{\rm DM}>M_{Z_p}$ as we have assumed,
dark matter can annihilate into a pair of $Z'$ via t-channel as shown in Figure \ref{Fig:fig5}(e).
Vertices related with these Feynman diagrams are given in Table \ref{table1} for $S_I$ annihilation.
As we have discussed above, $S_R$ can also plays the role of dark matter in the case of co-annihilation,
the relavant vertices are also given in Table \ref{table1}.

\begin{table}[htbp]
 \begin{tabular}{l|r}
 \hline
 Coupling  & Vertex Factor \\
 \hline
 ${\cal C}_{S_{I(R)} S_{I(R)} h_1}$     & $-i [ 2 v \lambda_{Hs} \cos\theta \pm 4 v_b \lambda_{ds}\sin\theta \mp 2 v_b \lambda_{SP} \sin\theta ]$\\
 \hline
 ${\cal C}_{S_{I(R)} S_{I(R)} h_2}$     & $-i [ 2 v \lambda_{Hs} \sin\theta \mp 4 v_b \lambda_{ds}\cos\theta \pm 2 v_b \lambda_{SP} \cos\theta ]$\\
 \hline
 ${\cal C}_{S_{I(R)} S_{I(R)} h_1 h_1}$ & $-i [ 2 \lambda_{Hs} \cos^2\theta  \mp 4 \lambda_{ds}\sin^2\theta + 2 \lambda_{SP} \sin^2\theta    ]$\\
 \hline
 ${\cal C}_{S_{I(R)} S_{I(R)} h_2 h_2}$ & $-i [ 2 \lambda_{SP} \cos^2\theta  \mp 4 \lambda_{ds}\cos^2\theta + 2 \lambda_{Hs} \sin^2\theta    ]$\\
 \hline
 ${\cal C}_{S_{I(R)} S_{I(R)} h_1 h_2}$ & $-i \cos\theta \sin\theta [ 2 \lambda_{Hs} \pm 4 \lambda_{ds} - 2\lambda_{SP} ]$\\
 \hline
 ${\cal C}_{S_I S_R Z'}$      & $\frac{2}{3}g_p(p_1-p_2)^{\mu}$\\
 \hline
 ${\cal C}_{S_{I(R)} S_{I(R)} Z'Z'}$      & $i 16 g_p^2$\\
 \hline
  \end{tabular}
  \caption{ Vertices related with $S_{I(R)}$ annihilation.}
  \label{table1}
\end{table}

In this work, we use {\cal Feynrules} \cite{Alloul:2013bka} to generate implemented code,
micrOmegas \cite{Belanger:2018ccd} to calculate the relic density and t3ps \cite{Maurer:2015gva} to scan the parameter space.
For simplicity, we assume $\lambda_s=0$ since the self interaction part of dark matter makes no contribution to the relic density.
Moreover, searches for the exotic Higgs at the LHC give upper limits of the mixing angle with $|\sin\theta| \leqslant 0.2 \sim 0.6$
depending on the heavy Higgs mass \cite{Accomando:2016sge,Robens:2015gla,Lopez-Val:2014jva}.
In addition, as we discussed above, $M_{Z_p}$ and $g_p$ are limited with current experiments. 
To sum up, there are 8 independent parameters in the model, 
and we take three Scenarios to estimate these parameters seprately for simplicity. 
In $Scenario\ A$, we consider $(g-2)_{\mu}$ anomaly as well as dark matter relic density constraint on the $M_{Z_p}-g_p$. 
In $Scenario\ B$, we discuss contribution of other parameters on the dark matter relic density. 
In $Scenario\ C$, we focus on the relic density constraint on the couplings $\lambda_{Hs}$, $\lambda_{SP}$ and $\lambda_{ds}$.

\subsubsection*{Scenario A}

\begin{figure}[htbp]
  \vspace{0.5cm}
  \hspace{-0.8cm}
  \includegraphics[scale=0.6]{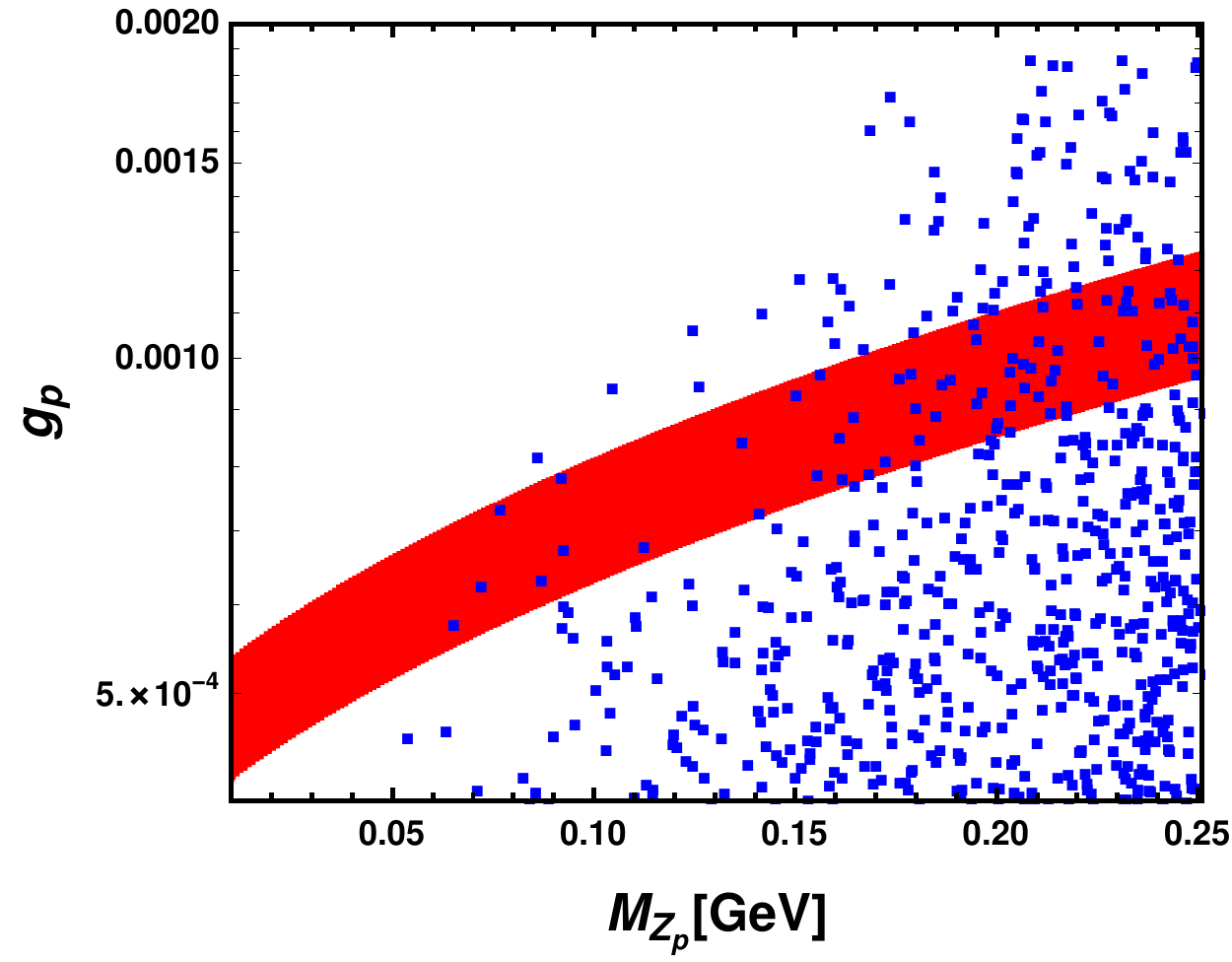}
  \caption{
  Allowed $M_{Z_p}-g_p$ region to satisfy relic density constriant [blue dots] and $(g-2)_{\mu}$ anomaly [red region]
  with $M_{Z_p}\subseteq [0.01,0.25]$ GeV and $g_p \subseteq [4 \times 10^{-4}, 0.002]$
  where we have set $\sin\theta=0.01$, $m_2\subseteq [0.2,2]$ TeV and $\lambda_{Hs, SP, ds} \subseteq [10^{-5},0.1]$, $m_0\subseteq [0,60]$ GeV.
  }
  \label{Fig:fig6}
\end{figure}

In this part, we focus on $(g-2)_{\mu}$ anomaly as well as dark matter relic density constraint on the $M_{Z_p}-g_p$, the inputs are set by Table \ref{tablea}.
 \begin{table}[htpb]
 \begin{tabular}{c|c}
 \hline
 Parameter &  value for inputs  \\
 \hline
 $m_2$ & $[0.2,2]$ TeV \\
 \hline
 $ \sin \theta $ & 0.001 \\
 \hline
 $\lambda_{Hs}$ &  $[10^{-5},0.1]$\\
 \hline
 $\lambda_{SP}$ &  $[10^{-5},0.1]$\\
 \hline
 $\lambda_{ds}$ & $ [10^{-5},0.1]$\\
 \hline
 $M_{Z_p}$ & $ [0.01,0.25]$ GeV\\
 \hline
 $g_p$ & $ [4 \times 10^{-4},0.002]$ \\
 \hline
 $m_0$ & $ [0 ,60]$ GeV \\
 \hline
  \end{tabular}
  \caption{ Values for the parameters as input to calculate dark matter relic density.}
   \label{tablea}
  \end{table}
According to Figure \ref{Fig:fig6}, we give the allowed region to satisfy relic density constriant (blue dots) and $(g-2)_{\mu}$ anomaly (red region)
with $M_{Z_p}\subseteq [0.01,0.25]$ GeV and $g_p \subseteq [4 \times 10^{-4},0.002]$
where we have set $\sin\theta=0.01$, $m_2\subseteq [0.2,2]$ TeV and $\lambda_{Hs, SP, ds} \subseteq [10^{-5},0.1]$.
Since we have chosen $m_2$ and $v_b=M_{Z_p}/2g_p$ as inputs,
these parameters should be limited by perturbative constraint as well as vacuum stability constriant,
which means $v_b=M_{Z_p}/2g_p$ should not be too small.
As we can see from  Figure \ref{Fig:fig6}, most of the blue points fall in the lower-right region,
and the upper-left region is excluded within the chosen parameter space.

\subsubsection*{Scenario B}

\begin{figure}[htbp]
  \includegraphics[scale=0.52]{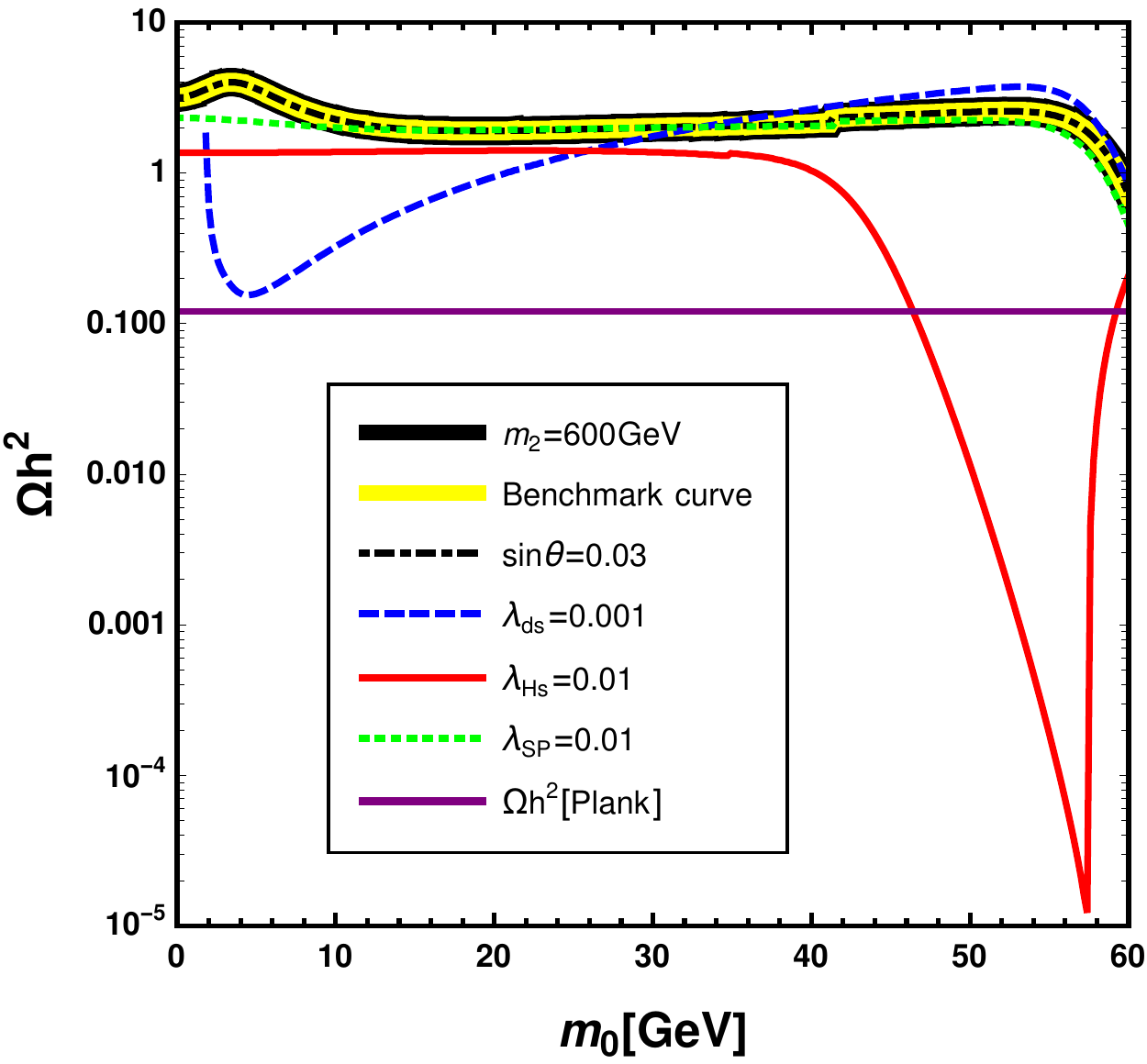}
  \includegraphics[scale=0.52]{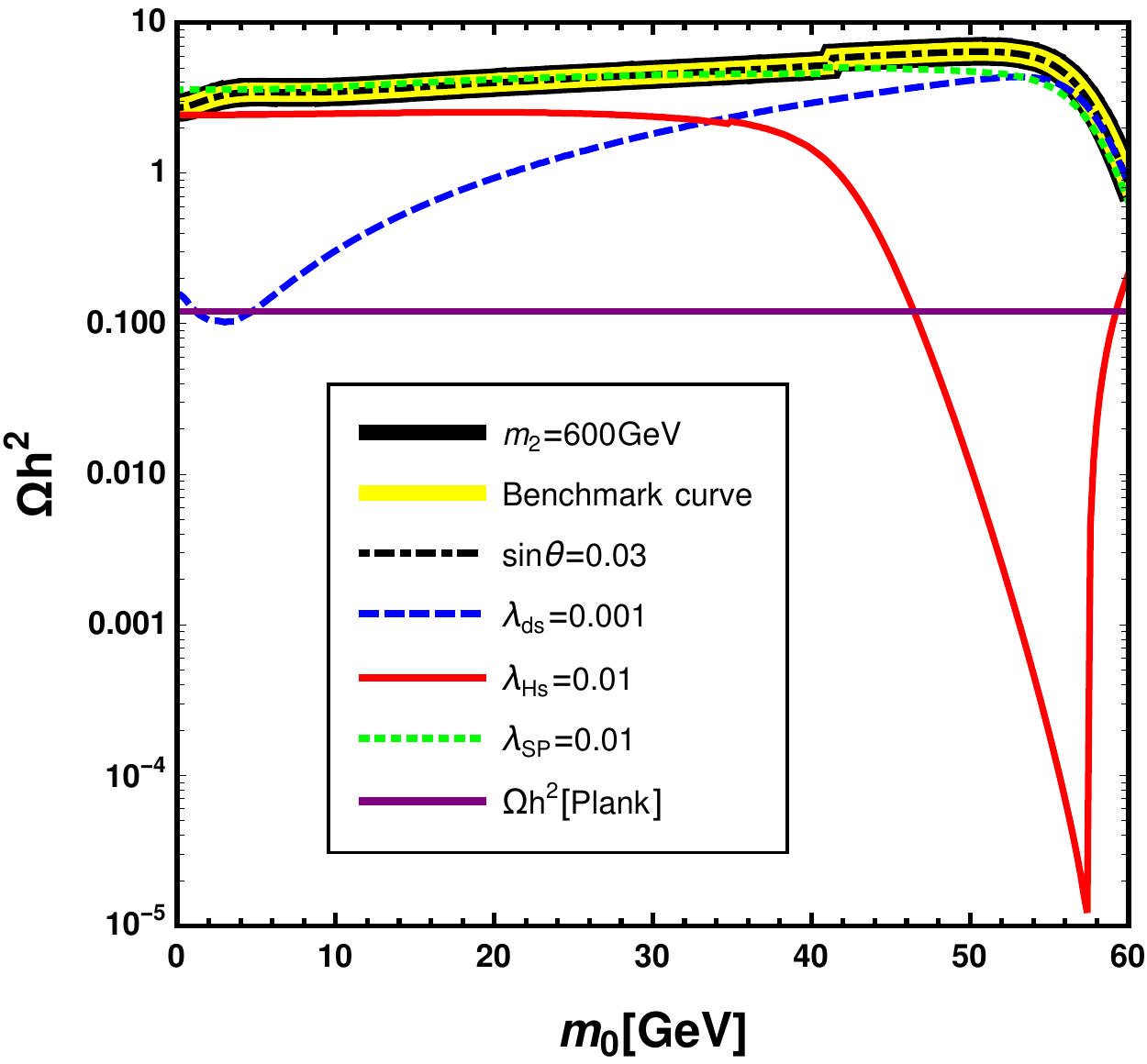}
  \caption{
  The value of Relic density as function of $m_0$. The purple solid curve is the observed relic density at experiment \cite{Aghanim:2018eyx}.
  Curves with different color corresponding to different parameter varies when $M_{Z_p}=0.2$ GeV, $g_p=0.001$ [first figure] and $M_{Z_p}=0.1$ GeV, $g_p=0.0007$ [second figure].}
  \label{Fig:fig7}
\end{figure}
In this part, we discuss contribution of other parameters to the dark matter relic density.
The results of relic density are shown in Figure \ref{Fig:fig7},
where curves with different color corresponding with one of the parameter varies. 
We set $0 \leq m_0 \leq 60\rm GeV$ and benchmark value for the parameters are given in Table \ref{table3}.
We fix $M_{Z_p}=0.2$ GeV and $g_p=0.001$ in Figure \ref{Fig:fig7}[first],
while in Figure \ref{Fig:fig7}[second] we fix $M_{Z_p}=0.1$ GeV and $g_p=0.0007$.
\begin{table}
 \begin{tabular}{c|c}
 \hline
 Parameter & benchmark value for inputs  \\
 \hline
 $m_2$ & 300 GeV \\
 \hline
 $ \sin\theta $ & 0.001 \\
 \hline
 $\lambda_{Hs}$ & $1 \times 10^{-4}$\\
 \hline
 $\lambda_{SP}$ & $1 \times 10^{-4}$\\
 \hline
 $\lambda_{ds}$ & $1 \times 10^{-4}$\\
 \hline
 $M_{Z_p}$ & 0.2 GeV \\
 \hline
 $g_p$ & 0.001\\
 \hline
  \end{tabular}
  \caption{Benchmark value for the parameters as input to calculate dark matter relic density.}
   \label{table3}
\end{table}

According to Figure \ref{Fig:fig7}[first], the purple solid curve is the observed relic density at experiments.
Curves corresponding to $m_2=600$ GeV and $\sin\theta=0.03$ almost coincide with the benchmark curve,
this is due to dark matter annihilation channels related with Higgs make little contribution to relic denisty in these cases.
Correspondingly, for the red line with $\lambda_{Hs}=0.01$, we have a resonance region at about $2m_{\rm DM}=m_1$,
where the relic density drops sharply, and intersect with the relic density constraint curve.
For $\lambda_{ds}=0.001$, see the blue dashed curve, the dark matter mass can be negative when $m_0$ takes small value,
so that the start point of the curve is not $m_0=0$.
For Figure \ref{Fig:fig7}[second] with $M_{Z_p}=0.1$ GeV and $g_p=0.0007$,
the curves are almost the same with those in Figure \ref{Fig:fig7}[first], 
and one of the typical difference is that the blue dashed curve corresponding to $\lambda_{ds}=0.001$ starts at $m_0=0$
and intersect with the relic density constraint curve.
According to these figures, we can reduce the inputs to $\lambda_{ds}$,
$\lambda_{SP}$, $\lambda_{Hs}$, $m_0$, $M_{Z_p}$ and $g_p$, 
since contribution of $m_2$ and $\sin\theta$ to relic density can be small.

\subsubsection*{Scenario C}

\begin{figure}[htbp]
  \centering
  \includegraphics[scale=0.3]{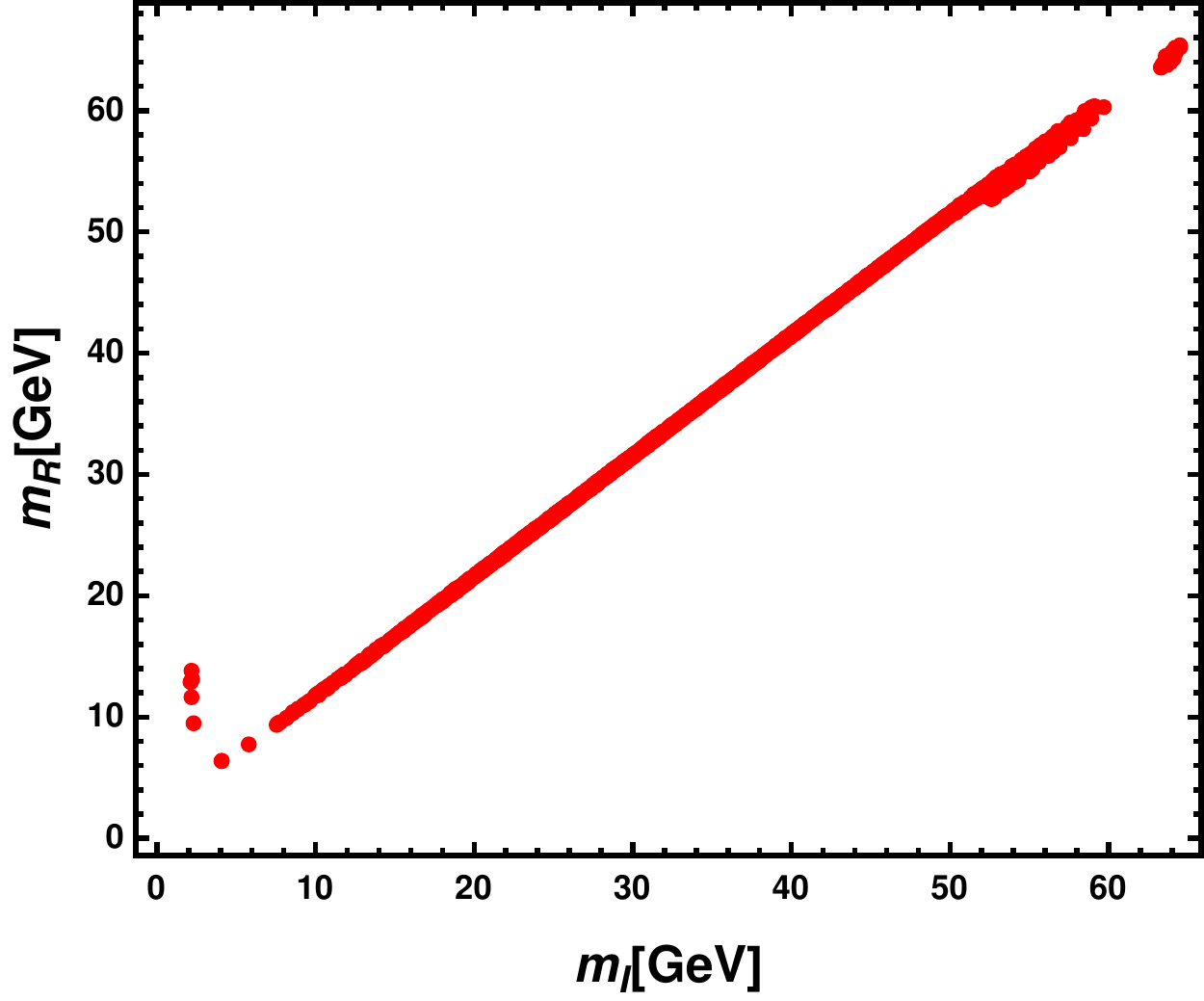}
  \includegraphics[scale=0.31]{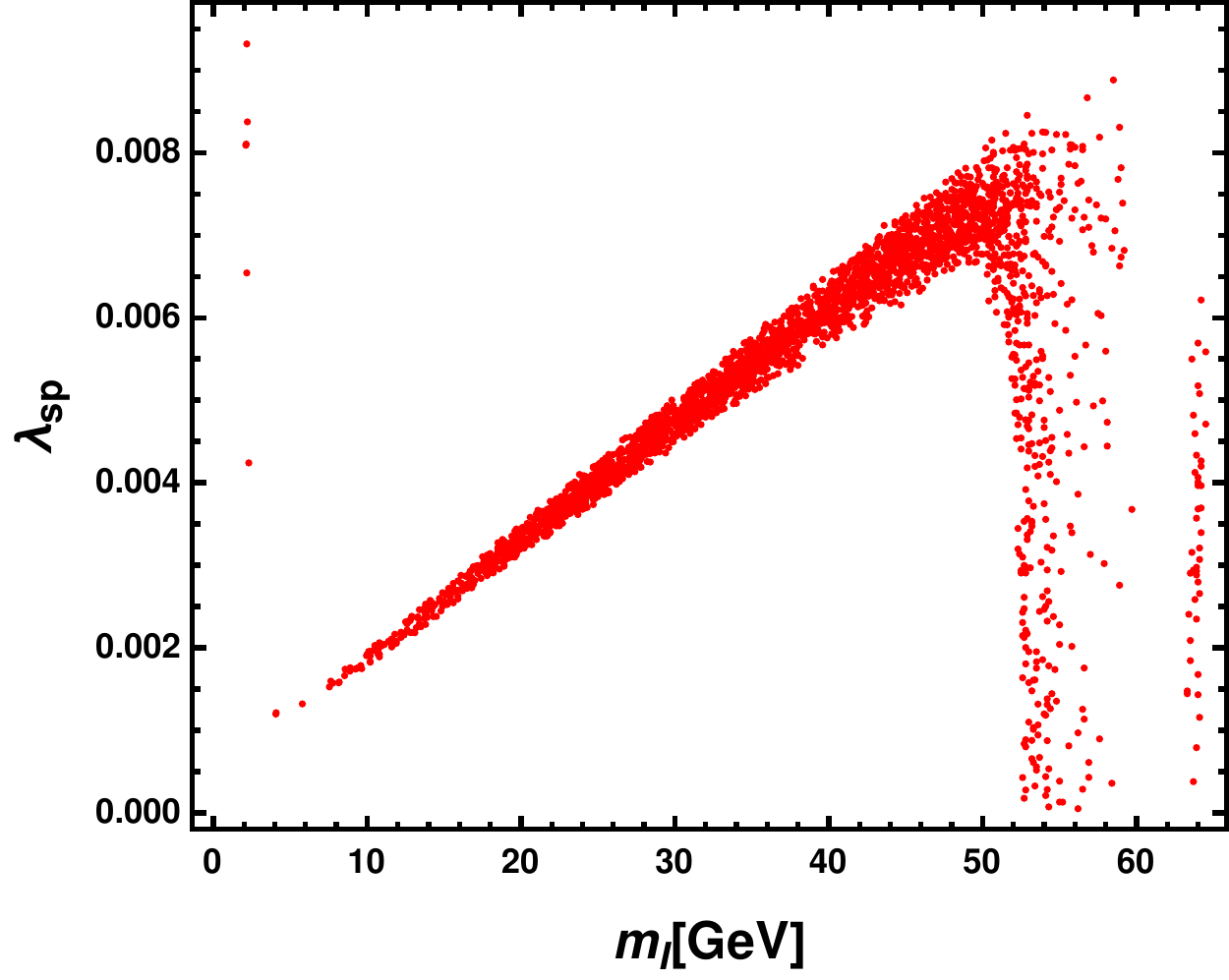}
  \includegraphics[scale=0.31]{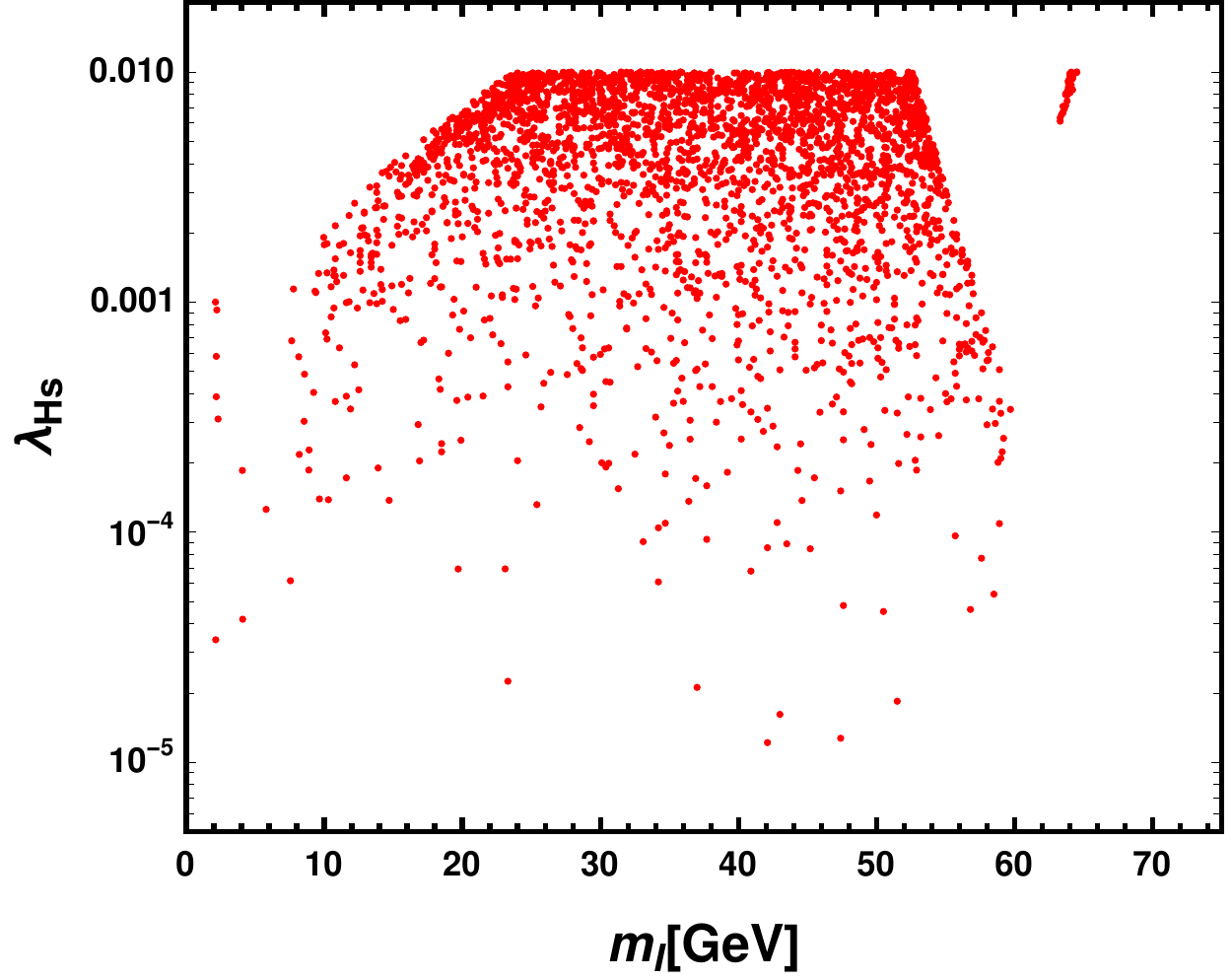}
  \includegraphics[scale=0.31]{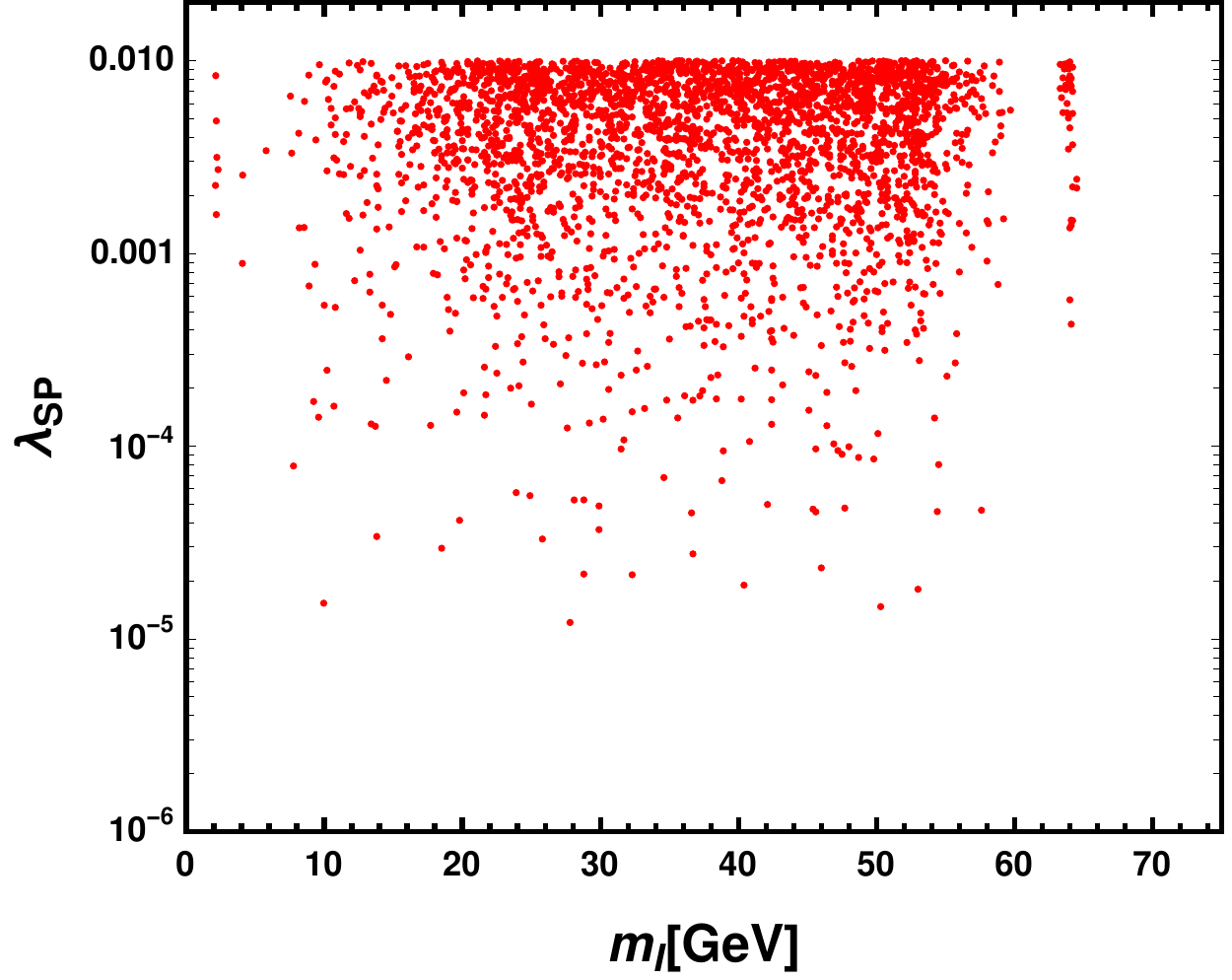}
  \caption{
  Scatter points satisfying relic density constraint with the X-axis being $S_I$ mass.
  Y-axis being $S_R$ mass in the first figure.
  The scan results of $\lambda_{ds}-m_I$, $\lambda_{Hs}-m_I$ and $\lambda_{SP}-m_I$ are in the second, third and fourth figure respectively.
  }
  \label{Fig:fig8}
\end{figure}

\begin{table}[htpb]
 \begin{tabular}{c|c}
 \hline
 Parameter &  value for inputs  \\
 \hline
 $m_2$ &  300 GeV \\
 \hline
 $ \sin \theta $ & 0.001 \\
 \hline
 $M_{Z_p}$ &  0.2 GeV\\
 \hline
 $g_p$ & 0.001 \\
 \hline
 $m_0$ & $ [0 ,60]$ GeV \\
 \hline
 $\lambda_{Hs}$ &  $[10^{-5},0.01]$\\
 \hline
 $\lambda_{SP}$ &  $[10^{-5},0.01]$\\
 \hline
 $\lambda_{ds}$ & $ [10^{-5},0.01]$\\
 \hline
  \end{tabular}
  \caption{ values for the parameters as input to scan.}
   \label{tablec}
\end{table}
In this part, we scan the parameter space to study the relic density constraint on the couplings. 
For simplicity, we fix $M_{Z_p}$, $g_p$, $\sin\theta$, $m_2$ and focus on $\lambda_{SP}$,
$\lambda_{ds}$ as well as $\lambda_{Hs}$. We set these parameters as in Table \ref{tablec}.
To avoid $m_{\rm DM}$ taking too large value, we have set the couplings $\lambda_{Hs, SP, ds}$ to be smaller than 0.01.
The results are given in Figure \ref{Fig:fig8}.
According to the first picture of Figure \ref{Fig:fig8},
the possible dark matter $S_I$ mass ranges from a few GeV to about $m_1/2$ as we discussed above, 
for $m_I $ bigger than about 10 GeV, we have $m_R \approx m_I$ and $\Delta \approx 0$ 
which means co-annihilation process can be domainate in the evolution of relic density 
where both $S_R$ and $S_I$ play the role of dark matter. 
For $S_I$ takes smaller value, $\Delta$ can be much large and only $S_I$ plays the role of dark matter, 
but such region is contrained stringently.
In addition, we give the scan result of $\lambda_{ds}-m_I$, $\lambda_{Hs}-m_I$
and $\lambda_{SP}-m_I$ in the other figures in Figure \ref{Fig:fig8} seperately.
As we can see from the second picture, the allowed value of $\lambda_{ds}$ is constrained 
within a narrow region when $m_I$ smaller than about 50 GeV,
increasing with $m_I$ increases, such result seemingly contraries to Eq.(\ref{eq8}),
where $m_I$ decreases with $\lambda_{ds}$ increase,
this means $\lambda_{Hs}$ and $\lambda_{SP}$ play more important role
determining dark matter mass under the relic density constraint.
For $m_I>50$ GeV, $\lambda_{ds}$ can take value within the whole range of $[1\times 10^{-5}, 0.0084]$.
For $\lambda_{Hs}$ and $\lambda_{SP}$ in the last two picture,
both can take value from $1 \times 10^{-5}$ to $0.01$,
and a certain number of the points fall in the region $[0,001, 0.01]$,
with a few points obtained at low mass region.

\subsection{Direct detection}\label{sec3.3}

Experiments related with dark matter direct detection such as CDMS II \cite{Ahmed:2009zw},
XENON100 \cite{Aprile:2017iyp}, XENON1T \cite{Aprile:2015uzo}, LUX \cite{Akerib:2013tjd}
have been searching for the signal of the interaction of DM with nucleon.
In our model, quarks do not couple with $Z'$ but only with two Higgs particles
and these interactions can be concluded by the scattering of dark matter particle off a SM fermion $f$
via the t-channel exchange of the two Higgs particles,
which is similar with the two singlet scalar model \cite{Abada:2011qb, Basak:2021tnj}.
The effective lagrangian for dark matter-quark elastic scattering can be given by :
\begin{eqnarray}
 {\cal L}_{q,eff}=-\frac{m_q}{2v}(\frac{{\cal C}_{h_1SS}\cos\theta}{m_1^2} + \frac{{\cal C}_{h_2SS}\sin\theta}{m_2^2})SS\bar{q}q ,
\end{eqnarray}
and effective lagrangian related with $S_I$ and $S_R$ can be given by:
\begin{eqnarray}
 {\cal L}_{q,eff}=\sum_{S=S_R,S_I} -\frac{m_q}{2v}(\frac{{\cal C}_{h_1SS}\cos\theta}{m_1^2}
                                                 + \frac{{\cal C}_{h_2SS}\sin\theta}{m_2^2})SS\bar{q}q .~
\end{eqnarray}
Furthermore, the effective lagrangian related with DM-nucleon elastic scattering can be given by:
\begin{eqnarray} \nonumber
 {\cal L}_{N,eff} &=& \sum_{S=S_R,S_I} \frac{m_N-\frac{7}{9}m_B}{v} \\
 && (\frac{{\cal C}_{h_1SS}\cos\theta}{m_1^2}+ \frac{{\cal C}_{h_2SS}\sin\theta}{m_2^2})SS\bar{N}N
\end{eqnarray}
where $m_N$ represents the nucleon mass and $m_B$ represents the baryon mass in the chiral limit \cite{He:2008qm}.
The total cross section for S-N elastic scattering can be given by:
\begin{eqnarray} \nonumber
 \sigma_{SN\rightarrow SN} &=&\frac{m^4_Nf_N^2}{4\pi (m_N+m_{\rm DM})^2} \\
  &\times& (\frac{{\cal C}_{h_1SS}\cos\theta}{m_1^2}+ \frac{{\cal C}_{h_2SS}\sin\theta}{m_2^2})^2,
\end{eqnarray}
where $f_N$ is the Higgs-nucleon Formfactor with $f_N=0.308(18)$
according to the phenomenological and lattice-QCD calculations \cite{Hoferichter:2017olk}
and $m_{\rm DM}$ corresponds to both $m_I$  and $m_R$.

\begin{figure}[htbp]
  \includegraphics[scale=0.65]{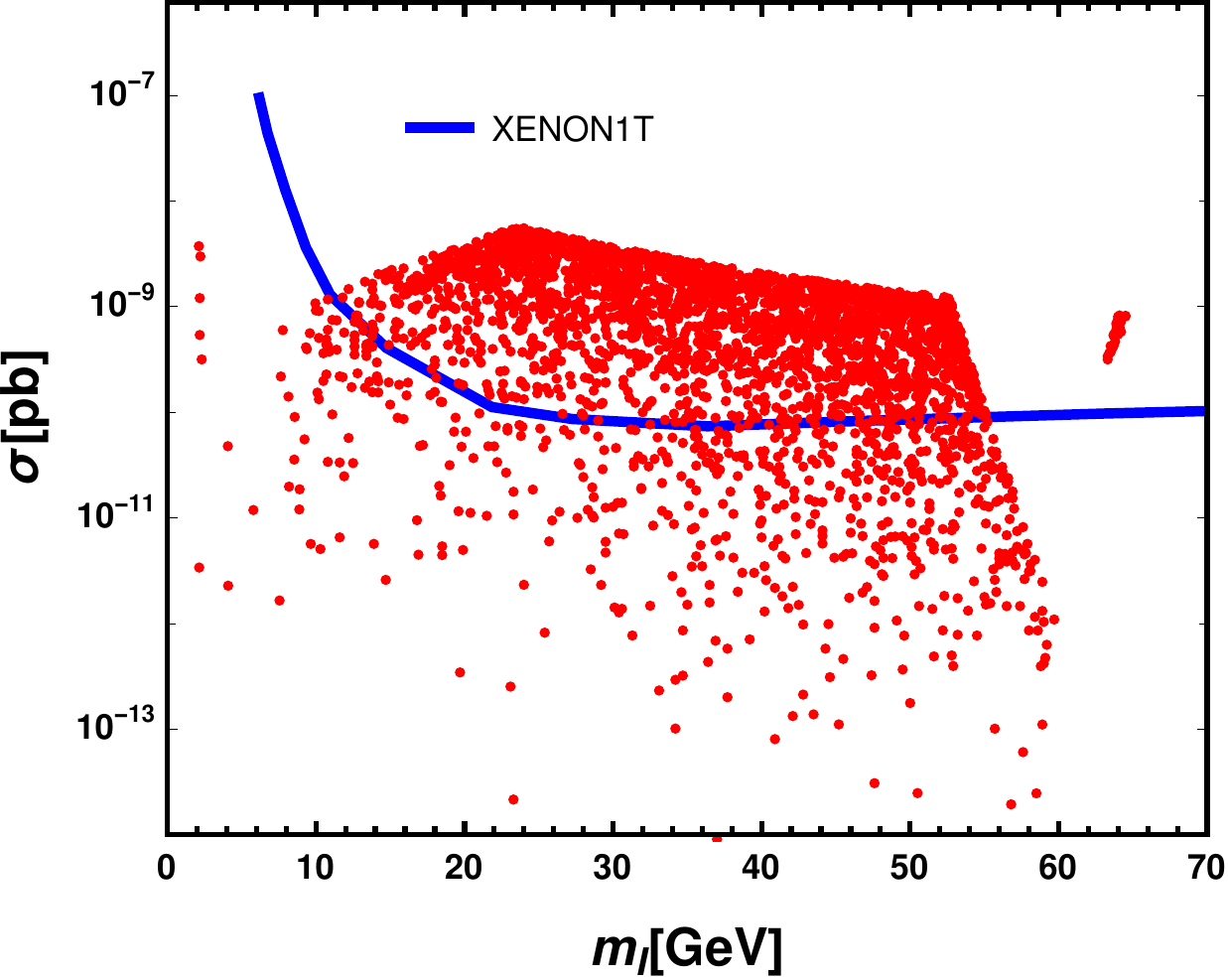}
  \caption{
  Spin indepedent cross section as a function of $m_{I}$,
  the red points represent result of the chosen parameter space satisfying relic density constraint,
  the blue dashed curve represents reslut of XENON1T \cite{Aprile:2015uzo}.}
 \label{Fig:fig9}
\end{figure}
We conisder the direct detection constraint on the chosen parameter space of the model
and the result is given in Figure \ref{Fig:fig9}, where the plot is drawn as a function of $m_I$.
The blue dashed curve in Figure \ref{Fig:fig9} represents reslut of XENON1T \cite{Aprile:2015uzo}
and the red points represent result of the chosen parameter space satisfying relic density constraint.
According to Figure \ref{Fig:fig9}, although a certain number of the red points fall above the region of direct detection constraint,
which means these points  are excluded by direct detection constraint. 
There remain points that mass ranging from a few GeV to about 60 GeV to both satisfy relic density constraint as well as direct detection constraint.
In other word, while direct detection can give stringent limit on the parameter space, 
dark matter mass can be from a few GeV to about 60 GeV within a viable parameter space satisfying relic density constraint in the model.

\section{Summary}\label{sec:so}

SM has achieved great success for its high accurancy to describle electroweak and strong interaction.
However, there remians problems such as neutrino mass and dark matter that SM can not explain.
In addition, recent results about muon anomalous magnetic moment $(g-2)_\mu$ brings new challenges to the SM.
The $4.2\sigma$ discrepancy between experiment and SM prediction seems to indicate new physics
behind $(g-2)_{\mu}$ anomaly and gives possible hints to the Beyond SM physics.
Among these models, a gauged $U(1)_{L_{\mu}-L{\tau}}$ model stands out for its simplicity
since the gauge anomaly cancellation can be accomplished without introducing extra fermions.
Related discussion about such model and experiment constrain the new guage boson $Z'$ mass at MeV scale,
which also satisfying the $(g-2)_{\mu}$ constraint.
In addition, one can also introduce new particles to the ${L_{\mu}-L_{\tau}}$ model
so that dark matter problem as well as neutrino mass problem can both be explained.

In this article, we consider a scalar dark matter model within the frame of a gauged ${L_{\mu}-L_{\tau}}$ model.
We introudce a new gauge boson $Z'$ as well as two scalar fields $S$ and $\Phi$ to the SM.
The lighter component of $S$ can play the role of dark matter
which is stablized by the residual $Z_2$ symmetry after spontaneously symmetry breaking.
In this work, we focus on dark matter and $(g-2)_{\mu}$ anomaly and ignore the neutrino mass problem. 
In the case of $m_I <m_1/2$ , $m_R <m_1/2$ and $M_{Z_p} <m_1/2$, 
we have new Higgs invisble channels which should be limited by current experiment result at LHC. 
A viable parameter space is considered to discuss the possibility of light dark matter as well as co-annihilation case.
We consider relic density constraint and find in the case of $m_I$ taking a few GeV, 
we can have the imaginary part $S_I$ of $S$ as light dark matter with $m_R$ much larger than $m_I$, 
but the  allowed parameter space is constrained stringently. However, with $m_I$ increasing, 
the allowed value for $m_I$ and $m_R$ is approximately equal so that we have $\Delta \approx 0$, 
and we come to so-called  co-annihilation process, where both $S_R$ and $S_I$ play the role of dark matter. 
At low mass region, relic density constraint limits $\lambda_{ds}$ stringently, 
which plays a significant role determing $m_I$ and $m_R$. 
In addition, direct detection has also been taken into consideration to constrain the chosen parameter space,
and the spin-independent dark matter nucleon elastic scattering cross section can give stringent constraint 
the viable parameter space. As we discussed above, we have shown this model can explain dark matter 
and $(g-2)_{\mu}$ anomaly at the same time in certain parameter space.

\acknowledgments
Hao Sun is supported by the National Natural Science Foundation of China (Grant No.12075043).

\bibliography{qx}
\bibliographystyle{apsrev}
\end{document}